\documentclass[twocolumn,aps,prd,preprintnumbers,showpacs,superscriptaddress,nofootinbib,amsmath,amssymb,floats,floatfix,showkeys,notitlepage,longbibliography]{revtex4-1}

\usepackage{comment}
\usepackage{graphicx}
\usepackage{subfigure}
\usepackage{palatino}
\usepackage[commandnameprefix=always]{changes}
\usepackage{hyperref}
\hypersetup{colorlinks=true,linkcolor=blue,urlcolor=blue,citecolor=blue}
\usepackage[toc,page]{appendix}
\usepackage[normalem]{ulem}

\usepackage{orcidlink}
\usepackage{lipsum}
\usepackage{graphicx}
\usepackage{subfigure}
\usepackage{palatino}
\usepackage{sans}
\usepackage{adjustbox}
\usepackage{latexsym}
\usepackage{amsmath}
\usepackage{amssymb}
\usepackage{amsfonts}
\usepackage{dcolumn}
\usepackage{bm}
\usepackage{tikz}
\usepackage{bigints}
\usepackage{array,tabularx,multirow,booktabs}
\usepackage[tracking=true]{microtype}
\SetTracking{}{500}
\SetTracking{encoding={*}, shape=sc}{40}
\UseRawInputEncoding 
\allowdisplaybreaks
\usepackage{adjustbox}
\usepackage{latexsym}
\usepackage{amsmath}
\usepackage{amssymb}
\usepackage{amsfonts}
\usepackage{dcolumn}
\usepackage{bm}
\usepackage{tikz}
\usepackage{bigints}
\usepackage{array,tabularx,multirow,booktabs}
\usepackage[tracking=true]{microtype}
\usepackage{color}

\def\1o2{{1\over2}}

\def\b{\beta}

\UseRawInputEncoding 
\allowdisplaybreaks

\begin{document} \sloppy

\title{Exact Regular Black Hole Solutions with de Sitter Cores and Hagedorn Fluid}

\author{Vitalii Vertogradov
\orcidlink{0000-0002-5096-7696}}
\email{vdvertogradov@gmail.com}
\affiliation{Physics department, Herzen state Pedagogical University of Russia,
48 Moika Emb., Saint Petersburg 191186, Russia.} 
\affiliation{SPB branch of SAO RAS, 65 Pulkovskoe Rd, Saint Petersburg
196140, Russia.}

\author{Ali \"Ovg\"un
\orcidlink{0000-0002-9889-342X}
}
\email{ali.ovgun@emu.edu.tr}
\affiliation{Physics Department, Eastern Mediterranean
University, Famagusta, 99628 North Cyprus, via Mersin 10, Turkiye.}

\begin{abstract}
In this paper, we present three exact solutions to the Einstein field equations, each illustrating different black hole models. The first solution introduces a black hole with a variable equation of state, $P = k(r)\rho$, which can represent both singular and regular black holes depending on the parameters $M_0$ and $w_0$. The second solution features a black hole with Hagedorn fluid, relevant to the late stages of black hole formation, and reveals similarities to the first solution by also describing both singular and regular black holes in a specific case. Furthermore, we investigate the shadows cast by these black hole solutions to constrain their parameters. Recognizing that real astrophysical black holes are dynamic, we developed a third, dynamical solution that addresses gravitational collapse and suggests the potential formation of naked singularities. This indicates that a black hole can transition from regular to singular and back to regular during its evolution.
\end{abstract}

\date{\today}



\maketitle

\section{Introduction}

Relativistic gravitational collapse is a fundamental concept in black hole physics, suggesting that a sufficiently massive star will inevitably form a black hole. This idea originates from the pioneering 1939 model by Oppenheimer, Snyder, and Datt \cite{Oppenheimer:1939ue,datt}. Their model describes a spherical, non-rotating, homogeneous cloud of matter composed of pressureless ‘dust’ particles. As this cloud collapses under its own gravity, it forms a black hole once its boundary crosses the Schwarzschild radius. Eventually, all the matter falls into a central singularity, concealed from distant observers.
Exploring the ultimate fate of gravitational collapse within Einstein’s theory of gravity is a highly active area in contemporary general relativity. Researchers are focusing on whether, and under what conditions, such collapse results in the formation of black holes. Additionally, they aim to identify physical collapse solutions that lead to naked singularities, which would challenge the cosmic censorship hypothesis. This hypothesis asserts that curvature singularities in asymptotically flat spacetimes are always hidden behind event horizons \cite{Penrose:1969pc,Misyura:2024fho,Konoplich:1999qq,Khlopov:1999ys,Khlopov:2008qy,Belotsky:2014kca,Dymnikova:2015yma}.

Building upon the pioneering work of Oppenheimer, Snyder, and Datt, numerous analytical studies of relativistic spherical collapse have explored various physical matter sources, such as dust and perfect fluids, both homogeneous and inhomogeneous. More recently, attention has shifted to models that incorporate corrections to general relativity at high densities. Static, spherically symmetric solutions to Einstein’s field equations, focusing primarily on isotropic fluids, are extensively detailed in the literature \cite{Stephani:2003tm,Delgaty:1998uy,Semiz:2008ny}. While isotropy is generally supported by observations, theoretical work suggests that local anisotropy may occur in high-density environments \cite{Herrera:1997plx}. Ruderman proposed that extremely compact objects could exhibit pressure anisotropy due to core densities exceeding nuclear density \textcolor{black}{($\approx10^{15}$ $g/cm^{3}$)}  (radial pressure $p_1$ and transverse pressure $p_2$) due to factors like solid cores and type-3A superfluids \cite{Ruderman:1972aj,Kim:2017hem}.

\textcolor{black}{Sakharov and Gliner \cite{Sakharov1966JETP...22..241S,Gliner1966JETP...22..378G} pioneered the study of regular black holes, proposing that essential singularities could be avoided by replacing the vacuum with a vacuum-like medium described by a de Sitter metric. Researchers such as Dymnikova, Gurevich, and Starobinsky later expanded upon this concept \cite{Dymnikova:1992ux,Gurevich1975Ap&SS..38...67G,Starobinsky:1979ty}. Bardeen \cite{Bardeen1968qtr..conf...87B} introduced the first practical regular black hole model, now known as the Bardeen black hole, by replacing the Schwarzschild black hole’s mass with a radius-dependent function.  The essential singularity in the Kretschmann scalar is eliminated by this innovation, which also gives the black hole a de Sitter core with positive Ricci curvature near its center \cite{Ansoldi:2008jw,Bambi:2023try,Lan:2023cvz}. In 1998, it was shown by Eloy Ayon-Beato and Alberto Garcia that regular black hole solutions can be constructed within General Relativity through the introduction of a nonlinear electrodynamic field \cite{Ayon-Beato:1998hmi,Ayon-Beato:1999qin}. Recently, Singh et al. demonstrated that regular black hole solutions can be achieved by coupling Einstein’s gravity with a nonlinear electrodynamics source \cite{Singh:2022xgi}. Moreover, an exact black hole solution for Einstein gravity coupled with nonlinear electrodynamics and a cloud of strings as the source was constructed, with its thermodynamical properties, quasinormal modes, shadow radius, and optical characteristics analyzed, revealing a second-order phase transition at a critical horizon radius in \cite{Singh:2022ycn}. An exact singular black hole solution was found by Sudhanshu et al. in the presence of nonlinear electrodynamics as the matter field source, surrounded by a cloud of strings in 4D AdS spacetime \cite{Sudhanshu:2024wqb}. } On the other hand, in 1965, Hagedorn proposed that for large masses \( m \), the hadron spectrum \( \rho(m) \) increases exponentially, \( \rho(m) \sim \exp(m/T_H) \), with \( T_H \) being the Hagedorn temperature \cite{Hagedorn:1965st}. This idea stemmed from the observation that at a certain point, adding more energy in proton-proton and proton-antiproton collisions ceases to increase the temperature of the resulting fireball and instead produces more particles, indicating a maximum achievable temperature \( T_H \) for a hadronic system \cite{Atick:1988si,Giddings:1989xe,Grignani:2001ik}. Moreover, The high-density Hagedorn phase of matter has been widely utilized in cosmology to explain the early stages of the Universe's evolution \cite{Maggiore:1997vw,Magueijo:2002pg,Bassett:2003ck}.

Since Einstein introduced the general relativity, we are entering a new era in gravitational physics, marked by two groundbreaking discoveries: first one was  in 2016 by the LIGO and VIRGO Collaborations which detected the first gravitational waves from the merger of two black holes \cite{LIGOScientific:2016aoc}, later including the coalescence of a black hole and a neutron star, while in 2019, the Event Horizon Telescope (EHT) Collaboration revealed the first-ever image of super-heated plasma around the supermassive object at the center of the M87 galaxy and later extended their observations to Sagittarius A* (Sgr A*) \cite{EventHorizonTelescope:2019dse,EventHorizonTelescope:2022wkp}. These findings provide strong evidence for the existence of supermassive black holes and the future advancements like will open doors to exploring alternative theories of gravity that go beyond our current understanding based on weaker gravitational fields \cite{Pantig:2022gih,Pantig:2022sjb,Kuang:2022xjp,Zakharov:2018awx,Zakharov:2018cbj,Virbhadra:1999nm,Claudel:2000yi,Virbhadra:2007kw,Virbhadra:2022iiy}. 

 This paper investigates the spherically symmetric gravitational collapse of matter in the Hagedorn phase, aiming to provide new insights into black hole solutions and the behavior of matter under extreme conditions. We begin by analyzing static models of regular black holes governed by a barotropic equation of state with an \(r\)-dependent coefficient. Next, we examine a static black hole with Hagedorn fluid and observe that, under certain conditions, this model resembles the first model with an \(r\)-dependent equation of state. Finally, we explore a dynamical model of gravitational collapse, which can result in the formation of either singular or regular black holes, depending on the initial conditions.

The gravitational collapse might lead to naked singularity formations~\cite{Joshi:1997de,Joshi:2008zz,Joshi:2013xoa,Dey:2019fpv}. This phenomenon has been investigated in many works~\cite{Vaidya:1951zz,Penrose:1964wq,Hawking:1970zqf,Vertogradov:2018ora,Shaikh:2019hbm,Firouzjaee:2021mwl,Vertogradov:2022zuo,Heydarzade:2023gmd,Vertogradov:2023uav,Kim:2017hem,Sajadi:2023ybm}. The process of regular black hole formation has not been widely investigated except for several works~\cite{Hayward:2005gi, Petrov:2023otl,Cai:2008mh,Culetu:2022otf,Mann:2021mnc,Simpson:2019cer, Joshi:2013xoa,Baccetti:2018qrp,Hossenfelder:2009fc,Ghosh:2000bc,Ghosh:2000ud,Ghosh:2008jca,Ghosh:2014pba,Hossenfelder:2009fc,Baccetti:2018qrp,Nasereldin:2023qph,Sharif:2015vya,Berezin:2016ubu,Babichev:2012sg,Malafarina:2016yuf,Harko:2003hs}. The final fate of the gravitational collapse of our model depends upon initial profile and might lead to naked singularity, singular or regular black hole formations. 

This work is organized as follows: In Sec. II, we present the exact solution for a regular black hole with a de Sitter core. This is then extended to include the Hagedorn fluid in Sec. III. In Sec. IV, we study the shadow cast by the newly obtained regular black holes with de Sitter cores and Hagedorn fluid, deriving constraints on the black hole parameters by comparing our models with the observed shadow of Sagittarius \(A^*\). In Sec. V, we explore a third, dynamical solution that examines gravitational collapse and suggests the possible formation of naked singularities, demonstrating how a black hole can transition between regular and singular states during its evolution. Finally, Sec. VI discusses the implications of the obtained results and outlines future research directions.

Throughout the paper, we will use the geometrized system of units $c=1=8\pi G$. Also we use signature $-+++$.

\section{Black hole with de Sitter core}
\textcolor{black}{In this section, we consider the toy model of regular black hole which, as we will find in next section, has connection to the particular case of a black hole with Hagedorn fluid.
In this study, the central singularity of a black hole is replaced with a de Sitter core, ensuring that all curvature scalars remain finite. This modification effectively resolves the singularity issue inherent in classical black hole models.}

Our objective is to explore a barotropic equation of state with a $r$-dependent coefficient,  $k\equiv k(r)$. To achieve this, we consider a general spherically symmetric, static spacetime expressed in the form:

\begin{equation} \label{eq:metric1}
d s^{2}=-f(r) d t^{2}+\frac{1}{f(r)} d r^{2}+r^{2} d \Omega^{2},
\end{equation}

where  $d\Omega^2=d\theta^2+\sin^2\theta d\varphi^2$ is the metric on unit two-sphere.  Then corresponding non-vanishing components of the Einstein field tensors ($G^{\mu}_{\nu}$) are calculated as:

\begin{eqnarray}
G_{t}^{t}=G_{r}^{r}=\frac{r f'(r)+f(r)-1}{r^2}, \\  G_{\theta}^{\theta}=G_{\phi}^{\phi}=\frac{f''(r)}{2}+\frac{f'(r)}{r},\end{eqnarray}
where the lapse function is chosen as
\begin{equation} \label{metric1}
f(r)=\left(1-\frac{2M(r)}{r}\right).
\end{equation}
Here $M(r)$ is an arbitrary function of $r$. 

The Einstein field equations for metric \eqref{metric1}, 

\begin{eqnarray}
G_{t}^{t}=G_{r}^{r}=-\frac{2 M'}{r^2}, \\  G_{\theta}^{\theta}=G_{\phi}^{\phi}=-\frac{M''}{r},\end{eqnarray} with the energy momentum tensor for an anisotropic
fluid is

\begin{eqnarray} \label{EQ7}
T_{\mu \nu}=\left(\rho+p_{2}\right) u_{\mu} u_{\nu}+\left(p_{1}-p_{2}\right) x_{\mu} x_{\nu}+p_{2} g_{\mu \nu}.\end{eqnarray}

\textcolor{black}{Here, $\rho$ represents the mass-energy density as measured by an observer comoving with the fluid, while $p_1$ and $p_2$ denote the radial and transverse pressures, respectively, with $p_2$ being the pressure in a direction perpendicular to the radial one. The prime ($'$) indicates differentiation with respect to $r$. Note that $u^{\mu}$ is the four-velocity of the fluid, and $x^{\nu}$ is a spacelike unit vector orthogonal to $u^{\mu}$, aligned along the angular directions}
(\(u^{\mu} u_{\mu}=-w^{\mu} w_{\mu}=
-1  \quad\)).

The radial and angular pressures
are assumed to be proportional to the density. Then energy momentum tensor is obtained as $T_{\theta}^{\theta}=T_{\phi}^{\phi}$ and 
$T_{t}^{t}=T_{r}^{r}$  and  $T^{\mu}{ }_{\nu}=\left(-\rho,-\rho,P,P \right)$ when the equation of state to be
\(p_{1}=-\rho\) and \(p_{2}=P\).

\textcolor{black}{Applying the above discussion to the spherically symmetric spacetime described by \eqref{eq:metric1}, we derive the corresponding Einstein equations as follows:}

\begin{eqnarray} \label{eq:einstein}
\rho=\frac{2M'}{r^2},\nonumber \\
P=-\frac{M''}{r}.
\end{eqnarray}

The system of differential equations \eqref{eq:einstein} consists of two equations and has three unknown functions $M$, $\rho$ and $P$. to close the system we need to introduce the equation of state, which we assume in the form
\begin{equation} \label{eq:state}
P=k(r)\rho.
\end{equation}
\textcolor{black}{The parameter $k(r)$ in the equation of state depends on the radial coordinate $r$. To satisfy the dominant energy conditions, we constrain its values to the range $k(r)$ $\in [-1, 1]$. Under this condition, and by considering the equations \eqref{eq:state} and \eqref{eq:einstein}, we arrive at a single differential equation of the following form:}
\begin{equation} \label{eq:prom}
M''r+2k(r)M'=0.
\end{equation}
In order to solve this equation we introduce new function $w(r)\equiv M'(r)$ and the equation \eqref{eq:prom} becomes
\begin{equation}
w'r+2kw=0,
\end{equation}
with the solution
\begin{equation}
w=w_0e^{-2\int \frac{k(r)}{r}dr}.
\end{equation}
Here $w_0$ is a constant of integration. Then the mass function $M(r)$ is given by
\begin{equation} \label{eq:mass1}
M(r)=\int \left[w_0e^{-2\int \frac{k(r)}{r}dr}\right ]dr+M_0,
\end{equation}
where $M_0$ another constant of an integration.

We consider the simple model in which $k(r)$ is the linear function of $r$ and we demand that at the star's surface $R$ it becomes zero and in the center it has a de Sitter core, i.e. $k(0)=-1$. 

\textcolor{black}{At the center of the de Sitter spacetime, the density reaches its maximum value, which is directly tied to the cosmological constant $\Lambda$. This aligns with the fundamental concept of linking the cosmological constant to the energy density arising from self interaction.\cite{Bambi:2023try}}

Thus, $k(r)$ has the form
\begin{equation} \label{eq:defk}
k(r)=\frac{r}{R}-1.
\end{equation}
Substituting \eqref{eq:defk} into \eqref{eq:mass1} gives after integration
\begin{equation} \label{eq:solution1}
M(r)=M_0-\frac{w_0}{2}\left(Rr^2+R^2r+\frac{R^3}{2}\right)e^{-\frac{2r}{R}}.
\end{equation}

\textcolor{black}{At  $r = 0$, M(r) is}
\begin{equation} \label{M_0}
 M(0) = M_0 - \frac{w_0 R^3}{4}. 
\end{equation}

The energy density $\rho$ and pressure are given by
\begin{eqnarray} \label{eq:reg_den}
\rho=2w_0e^{-\frac{2r}{R}},\nonumber \\
P=2w_0\left(\frac{r}{R}-1\right)e^{-\frac{2r}{R}}.
\end{eqnarray}
\textcolor{black}{From here, one can see that a constant of integration $w_0$ can be associated with  energy density in the center $\rho_0$ through the formula $w_0=\frac{\rho_0}{2}$.}
\textcolor{black}{To ensure a regular solution, it is essential to evaluate the scalar curvature invariants. The Ricci scalar $R_{icci}$ reads}
\begin{equation}\label{eq:ricci} 
R_{icci}=\frac{2 M''}{r}+\frac{4 M'}{r^{2}}
=\frac{\left(8 R-4 r\right) w_0 e^{-\frac{2 r}{R}}}{R}.
\end{equation}
\textcolor{black}{Ricci squared $S=R^{\mu \nu}R_{\mu \nu}$ is}
\begin{equation} \label{eq:ricci2}
S=\frac{2 \left(M''\right)^{2}}{r^{2}}+\frac{8 \left(M'\right)^{2}}{r^{4}}=\frac{8 w_0^{2} e^{-\frac{4 r}{R}} \left(2 R^{2}-2 R r+r^{2}\right)}{R^{2}}.
\end{equation}
\textcolor{black}{The Kretschmann scalar  is given by}

\begin{eqnarray}\label{eq:krech}
K=\frac{4 \left(M''\right)^{2}}{r^{2}}-\frac{16 M'' M'}{r^{3}}+\frac{16 M'' M}{r^{4}}+\frac{32 \left(M'\right)^{2}}{r^{4}} \notag\\-\frac{64 M' M}{r^{5}}+\frac{48 M^{2}}{r^{6}}.
\end{eqnarray}
\textcolor{black}{For the metric \eqref{eq:solution1} it has the form}

    \begin{eqnarray}
K =\frac{3 \left(\Theta\right) w_0^{2} e^{-\frac{4 r}{R}}-24 \left(\Gamma \right) R M_0}{R^{2} r^{6}}\end{eqnarray}
with
    \begin{eqnarray}
\Theta=R^{8}+4 r \,R^{7}+8 r^{2} R^{6}+\frac{32}{3} r^{3} R^{5}+12 r^{4} R^{4}\notag \\+\frac{32}{3} r^{5} R^{3}+\frac{32}{3} R^{2} r^{6}+\frac{16}{3} r^{8}
\end{eqnarray}
    \begin{eqnarray}
\Gamma=w_0 \left(R^{4}+2 R^{3} r +2 R^{2} r^{2}+\frac{4}{3} R \,r^{3}+\frac{4}{3} r^{4}\right) e^{-\frac{2 r}{R}} \notag\\-2 R M_0
\end{eqnarray}
\textcolor{black}{One sees that in order to have finite Kretschmann scalar, one must demand $M(0)=0$.}
In general, the spacetime \eqref{eq:solution1} has the singularity at $r=0$. Using the relation \ref{M_0}, we suggest this relation
\begin{equation} 
w_0=\frac{4M_0}{R^3},
\end{equation}
then the solution \eqref{eq:solution1} describes the non-singular black hole, where the finite Kretschmann scalar is:
 $r \to 0$ at the centre
\begin{equation}
\lim\limits_{r\rightarrow 0} K=\frac{512M_0^2}{3R^6}.
\end{equation}

 \textcolor{black}{Consequently, the curvature singularity is removed by the exponential factor, and the metric is interpreted as a non-singular black hole.}

\textcolor{black}{One should note that Ricci scalar $R_{icci}$ \eqref{eq:ricci} and squared Ricci $S$ \eqref{eq:ricci2} are finite regardless any additional conditions}
\begin{eqnarray} 
\lim\limits_{r\rightarrow 0}R_{icci}&=&8w_0,\nonumber \\
\lim\limits_{r\rightarrow 0}S&=&16w_0^2.
\end{eqnarray}

Finally, we arrive at the non-singular spacetime in the form
\begin{equation} \label{eq:regular_metric} 
ds^2=-f(r)dt^2+f^{-1}(r)dr^2+r^2d\Omega^2,
\end{equation}
where
\begin{equation}  \label{eq:lapse_regular1} 
f(r)=1-\frac{2M_0}{r}\left[1-\frac{2}{R^2}\left(r^2+rR+\frac{R^2}{2}\right)e^{-\frac{2r}{R}}\right].
\end{equation}

\textcolor{black}{At \(r \to \infty\) limit, the exponential term  $e^{-\frac{2r}{R}}$  decays to zero very rapidly, the remaining terms simplify as:
$f(r) \approx 1 - \frac{2M_0}{r}$, this solution behaves like Schwarzschild solution}
\begin{equation}
M(r) \approx M_0=const,~~r\rightarrow \infty,
\end{equation}
\textcolor{black}{and in the center it behaves like $\mathcal{O}(r^2)$. Thus,  $f(r)$  asymptotically approaches the Schwarzschild solution at large  $r$. On the other hand, as  $r \to 0$ :  $f(r) \to 1$  (finite, regular center).}

The weak energy condition states that \(T_{a b} t^{a} t^{b} \geq 0\) for all time like vectors \(t^{a}\) , i.e.,
for any observer, the local energy density must not be negative. Hence, the energy conditions
require \(\rho \geq 0\) and \(\rho+P_{i} \geq 0\), with \(P_{i}=-\rho-\frac{r}{2} \rho^{\prime}\).
\textcolor{black}{As one can see from \eqref{eq:reg_den}, the energy density is positive throughout spacetime if $w_0>0$, which we demand for regular black hole. The weak energy condition also requires $P+\rho\geq 0$ and from \eqref{eq:reg_den}, we have}
\begin{equation}
\rho+P=2w_0\frac{r}{R}e^{-2\frac{r}{R}}\geq 0,
\end{equation}
\textcolor{black}{that it is satisfied throughout spacetime for $w_0\geq 0$. Strong energy condition demands $\rho+P_1+2P_2\geq0$ and it is violated near the center but satisfied in the region $R\leq r < \infty$. The dominant energy condition is satisfied by construction because we demanded from the beginning that $-1\leq k(r)\leq 1$. This leads us to the region $r\leq 2R$ where dominant energy condition is held. However, we have constructed this model by demanding that $0\leq r \leq$ $R$.}

\section{Black hole with Hagedorn fluid}

The equation of state in ultra-dense region is based on the assumption that a whole host of baryonic resonant states arise at high densities. Hagedorn offered the model with equation of state \cite{Hagedorn:1965st,Malafarina:2016yuf,Harko:2003hs}
\begin{equation} \label{eq:hasdorf}
\bar{P}=P_0+\rho_0\ln \frac{\rho}{\rho_0},
\end{equation}
where $P_0$ and $\rho_0$ are constants.
\textcolor{black}{In this section, we write a line element in Eddington-Finkelstein coordinates $\{v, r, \theta, \varphi\}$ implying a subsequent transition to a dynamical model which is more convenient in these coordinates. The expression for the metric is given by}
\begin{equation} \label{hmetric} 
ds^2=-f(r)dv^2+2dv dr+r^2d\Omega^2,
\end{equation}
where, without loss of generality, we assume
\begin{equation} 
f(r)=1-\frac{2M(r)}{r}.
\end{equation}
The spacetime \eqref{hmetric} is supported, in general, with anisotropic energy-momentum tensor. However, one can calculate average pressure $\bar{P}$ as
\begin{equation} \label{eq:average}
\bar{p}=\frac{1}{3}\left(2P_2+P_1\right).
\end{equation}
However, the linearity and additivity of the Einstein tensor for the spacetime \eqref{hmetric} state that $G_0^0=G^1_1$ and $G^2_2=G^3_3$. It means that $P_1=-\rho$ and equation of state \eqref{eq:hasdorf} becomes
\begin{equation} \label{eq:hstate}
\frac{2}{3}P_2=P_0+\frac{1}{3}\rho+\rho_0 \ln \frac{\rho}{\rho_0}.
\end{equation}
The Einstein field equations for the metric \eqref{hmetric} are given by
\begin{eqnarray} \label{eq:heinstein}
\rho &=&\frac{2M'}{r^2},\nonumber \\
P_2&=&-\frac{M''}{r}.
\end{eqnarray}
In order to find the mass function $M(r)$ one should solve second order differential equation \eqref{eq:hstate} with \eqref{eq:heinstein}. To proceed, we introduce a new function
\begin{equation} \label{eq:h}
h(r)\equiv \frac{2M'}{r^2 \rho_0}.
\end{equation}
Then, we can obtain the following relation
\begin{equation}
-\frac{2M''}{3r}=-\frac{\rho_0}{3}rh'-\frac{2\rho_0}{3}h.
\end{equation}
Substituting it into \eqref{eq:hstate}, one obtains the following differential equation
\begin{equation} \label{eq:hdif2}
-rh'=3\alpha+3h+3\ln h.
\end{equation}

The formal solution is
\begin{equation} \label{eq:formal} 
r=\beta e^{-\zeta (h)},
\end{equation}
where $\beta$ an integration constant and
\begin{equation}
\zeta(h)\equiv \frac{1}{3}\int \frac{dh}{\alpha +h+\ln h},
\end{equation}
and $\alpha \equiv \frac{P_0}{\rho_0}\approx 0.25$.
The mass function then is given by
\begin{equation} \label{eq:hmass}
2M(r)=\int h r^2dr=-\frac{\beta^3}{3} \int \frac{h e^{-3h}dh}{\alpha+h+\ln h} +2 M_0,
\end{equation}
\textcolor{black}{where $M_0$ is another integration constant related to the black hole mass.}
When $\rho \rightarrow \rho_0$ the function $h$ becomes close to unity. So, doing in the integral \eqref{eq:formal} the transformation $w\equiv \ln h$ we can consider the solution \eqref{eq:formal} in power series of $\frac{r}{\beta}$, i.e.
\begin{equation}
\frac{r}{\beta}\approx 1- \frac{4}{15}w+\mathcal{O}\left(w^2\right).
\end{equation}
Remembering that $h=e^w$, we arrive at
\begin{equation}
h=e^{\frac{15}{4}\left(1-\frac{r}{\beta}\right)}.
\end{equation}
By using \label{eq:h} we arrive at the solution
\begin{equation} \label{eq:solution2}
M(r)=M_0-\frac{2\beta \rho_0}{15}\left(r^2+\frac{8 \beta r}{15}+\frac{32 \beta^2}{225}\right)e^{\frac{15}{4}\left(1-\frac{r}{\beta}\right)}.
\end{equation}

\textcolor{black}{This solution reminds \eqref{eq:solution1} and it also has regular center if }
\begin{equation}
M_0=\frac{64 \beta^3\rho_0}{3375} e^{\frac{15}{4}}.
\end{equation}
\textcolor{black}{Note, that this condition leads to $\lim\limits_{r\rightarrow 0} M(r)=0$. However, regardless this condition the energy density and pressure is a constant at $r\rightarrow 0$, that's why we need to impose only condition above. }

\textcolor{black}{Under our assumption, Hagedorn fluid behaves like the following equation of state}
\begin{equation} \label{eq:neod}
P=\left(-1+\frac{15r}{8\beta}\right)\rho,
\end{equation}
\textcolor{black}{and if we introduce $R$ by}
\begin{equation}
R\equiv \frac{8\beta}{15},
\end{equation}
\textcolor{black}{this solution becomes \eqref{eq:solution1}. Thus, the equation of state of Hagedorn fluid, under our assumption, transforms to equation of state \eqref{eq:neod}.}

The lapse function becomes
\begin{equation} \label{eq:solution2F(r)} 
f(r)=1-\frac{2 M_0}{r}+\frac{4 \beta  \rho_0 e^{\frac{15}{4} \left(1-\frac{r}{\beta }\right)} \left(32 \beta ^2+225 r^2+120 \beta  r\right)}{3375 r},
\end{equation}

and

\begin{equation} \label{eq:solution2F(r)smallbeta}
f(r)=1-\frac{2 M_0}{r}+e^{ \left(-\frac{15 r}{4 \beta }+\frac{15}{4}\right)}\left(\frac{4 r  \rho_0  \beta }{15}+\frac{32  \rho_0  \beta ^2}{225}\right)+O\left(\beta ^3\right).
\end{equation}

\textcolor{black}{For  $r \to \infty$, the exponential term  $e^{-\frac{15r}{4\beta}}$  dominates, and since it decays exponentially, the last term in  $f(r)$  becomes negligible. Thus:
$f(r) \approx 1 - \frac{2 M_0}{r}$, $\quad$ as $r \to \infty$. Thus,  $f(r)$  asymptotically approaches the Schwarzschild solution at large  $r$. On the other hand, as  $r \to 0$ :  $f(r) \to 1$  (finite, regular center).}

\textcolor{black}{To confirm the regularity of the solution \ref{eq:solution2F(r)}, we study the curvature invariants, including the Ricci square  \(S=R_{a b} R^{a b}\)   and the Kretschmann scalar  \(K=R_{a b c d} R^{a b c d}\), which are given by:}

\begin{widetext}
\begin{eqnarray}S=\frac{\rho_0^{2} \left(e^{\frac{15}{4}}\right)^{2} \left(e^{-\frac{15 r}{4 \beta}}\right)^{2} \left(128 \beta^{2}-240 r \beta +225 r^{2}\right)}{32 \beta^{2}},  \\ K=\frac{225 r^{2} \rho_0^{2} e^{\frac{15}{2}} e^{-\frac{15 r}{2 \beta}}}{16 \beta^{2}}+8 \rho_0^{2} e^{\frac{15}{2}} e^{-\frac{15 r}{2 \beta}}+\frac{64 \beta  \rho_0^{2} e^{\frac{15}{2}} e^{-\frac{15 r}{2 \beta}}}{15 r}+\frac{64 \beta^{2} \rho_0^{2} e^{\frac{15}{2}} e^{-\frac{15 r}{2 \beta}}}{25 r^{2}}+\frac{4096 \beta^{3} \rho_0^{2} e^{\frac{15}{2}} e^{-\frac{15 r}{2 \beta}}}{3375 r^{3}}+\frac{8192 \beta^{4} \rho_0^{2} e^{\frac{15}{2}} e^{-\frac{15 r}{2 \beta}}}{16875 r^{4}} \notag \\+\frac{32768 \beta^{5} \rho_0^{2} e^{\frac{15}{2}} e^{-\frac{15 r}{2 \beta}}}{253125 r^{5}}+\frac{65536 \beta^{6} \rho_0^{2} e^{\frac{15}{2}} e^{-\frac{15 r}{2 \beta}}}{3796875 r^{6}}-\frac{30 \mathit{M_0} e^{\frac{15}{4}} e^{-\frac{15 r}{4 \beta}} \rho_0}{\beta  r^{2}}-\frac{16 \mathit{M_0} e^{\frac{15}{4}} e^{-\frac{15 r}{4 \beta}} \rho_0}{r^{3}} -\frac{64 \beta  \mathit{M_0} e^{\frac{15}{4}} e^{-\frac{15 r}{4 \beta}} \rho_0}{5 r^{4}}\notag \\-\frac{512 \beta^{2} \mathit{M_0} e^{\frac{15}{4}} e^{-\frac{15 r}{4 \beta}} \rho_0}{75 r^{5}}-\frac{2048 \beta^{3} \mathit{M_0} {\mathrm e}^{\frac{15}{4}} e^{-\frac{15 r}{4 \beta}} \rho_0}{1125 r^{6}}+\frac{48 \mathit{M_0}^{2}}{r^{6}}.\end{eqnarray}\end{widetext}
and 

the Ricci scalar $R_{icci}$ reads
\begin{equation}\label{eq:ricci2}
R_{icci}=\frac{\left(16 \beta -15 r\right) e^{-\frac{15 r}{4 \beta}} e^{\frac{15}{4}} \rho_0}{4 \beta}.
\end{equation}

Note that at asymptotical limits,
\begin{eqnarray} 
\lim\limits_{r\rightarrow 0}R_{icci}&=&4 \,e^{\frac{15}{4}} \rho_0 ,\nonumber \\
\lim\limits_{r\rightarrow 0}S&=&4 e^{\frac{15}{2}} \rho_0^{2}\\ \lim\limits_{r\rightarrow 0}K&=&\frac{8 \rho_0^{2} e^{\frac{15}{2}}}{3} .
\end{eqnarray}


\section{Parameter constraints from shadows of black hole}

In this section, we explore constraints on the parameter $R$ by comparing the shadow of a black hole, as defined by Eq.\eqref{eq:lapse_regular1}, with the experimental image of Sagittarius $A^*$ obtained by the Event Horizon Telescope Collaboration (EHT)\cite{EventHorizonTelescope:2022wkp}. The shadow size provides direct insights into the spacetime geometry and the physical parameters defining the black hole.

First of all, we should note that the visible angular size of a Sagittarius $A^*$ described by Schwarzschild spacetime is \textcolor{black}{$\sim 53$ micro-arc-seconds ($\mu as$)} as which is larger than obtained image. Thus, the parameter $R$ should decrease the angular size of a shadow. 

Here, we use the method elaborated in the paper~\cite{Vertogradov:2024qpf}, which states that one can consider the lapse function $f$ in the form \begin{equation}f(r)=\left(1-\frac{2M_0}{r}\right)e^{\alpha g(r)}. \end{equation} Here $M_0$ is a mass of a black hole and $\alpha g(r)$ minimal geometrical deformation of the Schwarzschild spacetime. Then, we consider the radius  of a photon sphere $r_{ph}$ as
\begin{equation}r_{ph}=3M_0+\alpha r_1.\end{equation}
\textcolor{black}{Here, we assume small deviations from Schwarzschild spacetime and radius of a photon sphere decreases and increases in comparison with Schwarzschild case depending on the sign of $r_1$. In order to proceed we use method elaborated in the paper~\cite{Vertogradov:2024qpf}. The spacetime in general case is given by}
\begin{align}
ds^2 &= -\left(1-\frac{2M}{r}\right)e^{\alpha g(r)}dt^2 
+ \left(1-\frac{2M}{r}\right)^{-1}e^{-\alpha g(r)}dr^2 \notag\\
&\quad + r^2 d\Omega^2.
\end{align}

In the paper~\cite{Vertogradov:2024qpf} it has been proven that the visible angular size of a shadow decreases if $\alpha g(3M_0)>0$.In the case of the regular black hole \eqref{eq:lapse_regular1}, the minimal geometrical deformation $\alpha g(r)$ is given by \begin{equation}\alpha g(r)=\ln |\frac{r-2M_0\left(2X^2+2X+1\right)e^{-2X}}{r-2M_0}|.\end{equation}
Where we have denoted $X\equiv \frac{r}{R}$. In order to consider $\alpha g(3M_0)>0$, one should prove that the following inequality is held\begin{equation} \label{eq:picture1}\left(\frac{18M_0^2}{R^2}+\frac{6M_0}{R}+1\right)e^{-\frac{6M_0}{R}}<1.\end{equation}
This inequality is held for $R \in [0, 3M_0)$\footnote{Remember that in our model $P(R)=0$ and this should be at the radius $R<r_{ph}^{Schwarzschild}=3M_0$, otherwise there is no a shadow.} 
In order to find constraints, one should find such $R$ and $r_{ph}$ that the visible angle $\omega$, which is given by

\begin{equation}\sin^2\omega=\frac{b^2(r_{ph})}{b^2(r_o)},\end{equation}corresponds to obtained image. \textcolor{black}{Here $r_o$ is the distance from the Earth to Sagittarius $A^*$ and $b$ is the impact parameter.} For this purpose, one should solve the following system of equations \begin{eqnarray}\sin^2\omega&=&\frac{b^2(r_{ph})}{b^2(r_o)},\nonumber \\b(r_{ph})&=&\frac{r_{ph}}{\sqrt{f(r_{ph})}},\nonumber \\f'(r_{ph})r_{ph}&=&2f(r_{ph}).\end{eqnarray}
Here $f(r)$ is given in \eqref{eq:lapse_regular1} and the last equation is given by
\begin{equation}M_0\left(-4X^3_{ph}+6X^2_{ph}+6X_{ph}+1\right)e^{-2X_{ph}}-r_{ph}=0,\end{equation}where\begin{equation}X_{ph}\equiv \frac{r_{ph}}{R}.\end{equation}

\begin{figure}[htp]
   \centering
          \includegraphics[scale=0.6]{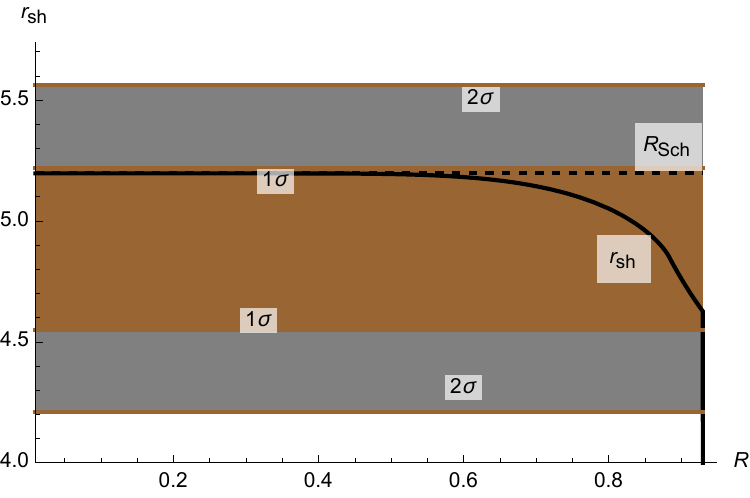}
    \caption{Constraints of black hole with de Sitter core from the Event Horizon Telescope horizon-scale image of Sagittarius A* at $1\sigma$\cite{Vagnozzi:2022moj}, after averaging the Keck and VLTI mass-to-distance ratio priors for the same with $M_0=1$, and varying $R$ .}
    \label{fig:shadow1}
\end{figure}

\textcolor{black}{The Fig. \ref{fig:shadow1} presents constraints on the shadow radius $(r_{\text{sh}})$ of a black hole with a de Sitter core as a function of the parameter $R$, derived from the Event Horizon Telescope (EHT) horizon-scale image of Sagittarius A* at different confidence levels ($1\sigma$ and $2\sigma$).}

The shadow radius of the black hole with a de Sitter core decreases as the value of \( R \) increases shown in \ref{fig:shadow1}. Moreover Fig. \ref{fig:shadow1} displays the upper limits of \(R\) based on EHT observational results for Sgr A*. The \(68\%\) confidence level (C.L.) \cite{Vagnozzi:2022moj} indicates that the upper limit for \(R\) is \(0.9\).

\begin{table}[ht!]
    \centering
    \begin{tabular}{ |p{1cm}||p{2cm}|p{2cm}| }
    \hline
        $R$ &  $r_{ph}$ &  $r_{sh}$ \\ [0.5ex] 
\hline
0.1 & 3. & 5.19615 \\
\hline
0.3 & 2.99999 & 5.19615 \\
\hline
0.5 & 2.99297 & 5.1934 \\
\hline
0.7 & 2.8941 & 5.14318 \\
\hline
0.9 & 2.17338 & 4.76121 \\
\hline
    \end{tabular}
    \caption{Effects of the parameter $R$ on the shadow of the black hole with de Sitter core $r_{sh}$ and photon sphere $r_{ph}$ for fixed $M_0=1$.}
    \label{table1}
\end{table}

\textcolor{black}{The data presented in Table \ref{table1} explores the influence of the parameter \( R \) on the shadow radius (\( r_{sh} \)) and photon sphere radius (\( r_{ph} \)) of a black hole with a de Sitter core, for a fixed \( M_0 = 1 \). Both \( r_{ph} \) and \( r_{sh} \) exhibit a decreasing trend as \( R \) increases, with \( r_{ph} \) showing a sharper decline, particularly at larger values of \( R \). For small \( R \), \( r_{ph} \approx 3 \) and \( r_{sh} \approx 5.196 \) remain relatively stable and similar to Schwarzschild case; however, at \( R = 0.9 \), they significantly decrease to \( 2.17338 \) and \( 4.76121 \), respectively. This behavior suggests that \( R \) has a nonlinear effect on spacetime geometry, reducing both the photon sphere and the apparent shadow size. While \( r_{sh} \) is consistently larger than \( r_{ph} \) due to gravitational lensing, their proportional decrease highlights a strong correlation influenced by \( R \). These results emphasize \( R \)’s critical role in determining observable black hole characteristics and its potential to differentiate black holes with de Sitter cores from classical models.}

\begin{figure}[htp]
   \centering
          \includegraphics[scale=0.6]{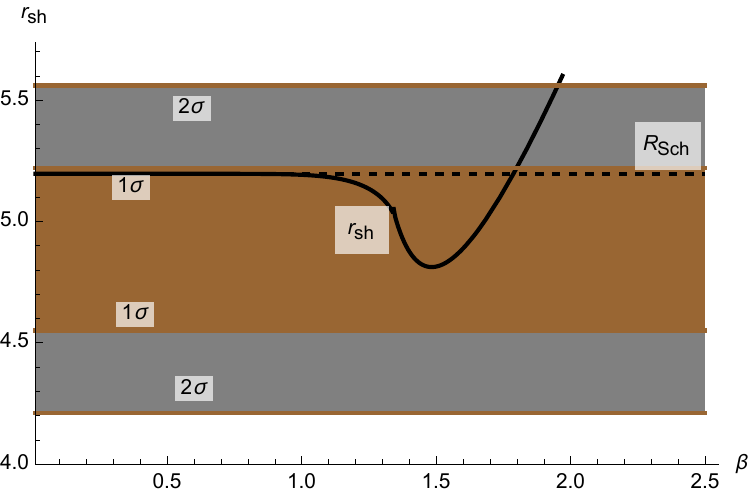}
    \caption{Constraints of black hole with Hagedorn fluid from the Event Horizon Telescope horizon-scale image of Sagittarius A* at $1\sigma$\cite{Vagnozzi:2022moj}, after averaging the Keck and VLTI mass-to-distance ratio priors for the same with $M_0=1$, and varying $\beta$ .}
    \label{fig:shadow2}
\end{figure}

Afterwards, we plot the Fig. \ref{fig:shadow2} that the radius of the black hole with Hagedorn fluid's shadow (\(r_{sh}\)) is computed for varying values of $\beta$. The shadow radius of the black hole with a Hagedorn fluid first decreases as the value of $\beta$ increases, then shadow radius increases as  shown in \ref{fig:shadow2}. Moreover Fig. \ref{fig:shadow2} displays the upper limits of $\beta$ based on EHT observational results for Sgr A*. The \(68\%\) confidence level (C.L.) \cite{Vagnozzi:2022moj} indicates that the upper limit for $\beta$ is \(1.8\).

\begin{table}[ht!]
    \centering
    \begin{tabular}{ |p{1cm}||p{2cm}|p{2cm}| }
    \hline
        $\beta$ &  $r_{ph}$ &  $r_{sh}$ \\ [0.5ex] 
\hline
0.4 & 3. & 5.19615 \\
\hline
0.8 & 2.99944 & 5.19596 \\
\hline
1.2 & 2.91852 & 5.15827 \\
\hline
1.6 & 3.12694 & 4.88834 \\
\hline
2. & 4.29899 & 5.67358 \\
\hline
    \end{tabular}
    \caption{Effects of the parameter $\beta$ on the shadow of the black hole with Hagedorn fluid $r_{sh}$ and photon sphere $r_{ph}$ for fixed $M_0=1$ and $\rho_0=1$.}
    \label{table2}
\end{table}

\textcolor{black}{Table \ref{table2} analyzes the impact of the parameter \( \beta \) on the photon sphere radius (\( r_{ph} \)) and shadow radius (\( r_{sh} \)) of a black hole with a Hagedorn fluid core, for fixed \( M_0 = 1 \) and \( \rho_0 = 1 \). Unlike a monotonic trend, \( r_{ph} \) and \( r_{sh} \) demonstrate non-linear behavior with varying \( \beta \). At lower values (\( \beta = 0.4 \) and \( \beta = 0.8 \)), both \( r_{ph} \) and \( r_{sh} \) remain close to their initial values (similarly the Schwarzschild case) of \( 3 \) and \( 5.19615 \), respectively. For moderate \( \beta \) (\( \beta = 1.2 \)), \( r_{ph} \) slightly decreases to \( 2.91852 \) while \( r_{sh} \) begins to drop to \( 5.15827 \). At \( \beta = 1.6 \), \( r_{ph} \) surprisingly increases to \( 3.12694 \) while \( r_{sh} \) drops significantly to \( 4.88834 \), indicating a shift in the geometry of spacetime. At high \( \beta = 2.0 \), both values increase, with \( r_{ph} \) reaching \( 4.29899 \) and \( r_{sh} \) increasing to \( 5.67358 \). These results reveal that \( \beta \) introduces complex dynamics in the spacetime structure, influencing both the photon sphere and shadow radius non-linearly. This suggests a nuanced interplay between \( \beta \) and the Hagedorn fluid parameters, making \( \beta \) a critical factor in determining observable black hole characteristics.}

\begin{table}[ht!]
    \centering
    \begin{tabular}{ |p{1cm}||p{2cm}|p{2cm}| }
    \hline
        $\rho_0$ &  $r_{ph}$ &  $r_{sh}$ \\ [0.5ex] 
\hline
0.1 & 2.99901 & 5.19574 \\
\hline
0.5 & 2.99501 & 5.19407 \\
\hline
0.9 & 2.99091 & 5.19238 \\
\hline
1.3 & 2.98671 & 5.19066 \\
\hline
1.7 & 2.98241 & 5.18891 \\
\hline
2.1 & 2.978 & 5.18713 \\
\hline
2.5 & 2.97347 & 5.18532 \\
\hline
    \end{tabular}
    \caption{Effects of the parameter $\rho_0$ on the shadow of the black hole with Hagedorn fluid $r_{sh}$ and photon sphere $r_{ph}$ for fixed $M_0=1$ and $\beta=1$.}
    \label{table3}
\end{table}

\textcolor{black}{Table \ref{table3} investigates the influence of the parameter \( \rho_0 \) on the photon sphere radius (\( r_{ph} \)) and shadow radius (\( r_{sh} \)) of a black hole with a Hagedorn fluid core, for fixed \( M_0 = 1 \) and \( \beta = 1 \). The results demonstrate a gradual, nearly linear decrease in both \( r_{ph} \) and \( r_{sh} \) as \( \rho_0 \) increases from 0.1 to 2.5. At \( \rho_0 = 0.1 \), the photon sphere radius is \( r_{ph} = 2.99901 \), and the shadow radius is \( r_{sh} = 5.19574 \). These values slightly decrease to \( r_{ph} = 2.97347 \) and \( r_{sh} = 5.18532 \) for \( \rho_0 = 2.5 \). The consistent trend suggests that higher \( \rho_0 \) values, corresponding to greater densities of the Hagedorn fluid, lead to a contraction of both the photon sphere and shadow radius. This behavior highlights the subtle impact of \( \rho_0 \) on the spacetime geometry around the black hole, suggesting a strong correlation between the fluid density and the observable black hole properties. Despite the gradual decrease, the values remain close across the range, indicating that \( \rho_0 \) exerts a steady but moderate influence compared to parameters like \( \beta \).}

\textcolor{black}{Deviations in  $r_{ph}$  directly modify the shadow size. Parameters  $R$, $\beta$ and $\rho_0$ signify physical deformations from general relativity. Constraints on these parameters help test the validity of extended gravity theories.}

\section{Formation of regular black hole and naked singularities}

As shown in the previous sections, the solutions \eqref{eq:solution1} with the equation of state \(P = k(r)\rho\) (where \(k(r) = A + Br\)) and \eqref{eq:solution2} with the Hagedorn fluid exhibit similar behavior under certain conditions. In this section, we study a gravitational collapse model governed by the equation of state \eqref{eq:defk}, within the spacetime described in Eddington-Finkelstein coordinates \(\{v, r, \theta, \varphi\}\):

\begin{equation} \label{eq:metcol}
ds^2=-\left(1-\frac{2M(v,r)}{r}\right)dv^2+2dvdr+r^2d\Omega^2,
\end{equation}
here, $M(v,r)$ is the mass function and $v$ advanced Eddington time. The spacetime \eqref{eq:metcol} is supported by combination of two energy-momentum tensors of type-I and II describing null dust and null fluid respectively. This energy-momentum tensor can be written as
\begin{equation} \label{emtgeneral}
T^{tot}_{\mu \nu}=T^{ND}_{\mu \nu}+T^{NS}_{\mu \nu},
\end{equation}
where $T^{ND}_{\mu \nu}$ is the energy-momentum tensor of null dust 
\begin{equation} \label{eq:emtvaidya} 
T^{(ND)}_{\mu \nu}=\sigma(v,r)l_\mu l_\nu,
\end{equation}
and $T^{NS}_{\mu \nu}$ represents null fluid
\begin{equation} \label{eq:emtgeneralized}
T^{(NS)}_{\mu \nu}=(\rho+P)(l_\mu n_\nu+l_\nu n_\mu)+Pg_{\mu \nu},
\end{equation}
where $l^\mu$ and $n^\mu$ are two null vectors with properties
\begin{equation}
l_\mu l^\mu=n_\mu n^\mu=0,~~ n_\mu l^\mu=-1,
\end{equation}
and have the form
\begin{eqnarray}
l_\mu&=&\delta^0_\mu,\nonumber \\
n_{\mu}&=&\frac{1}{2} \left (1-\frac{2M}{r} \right )\delta^0_{\mu}-\delta^1_{\mu}.
\end{eqnarray}

Here, \(\sigma\) represents the energy density of the null dust, while \(\rho\) and \(P\) denote the energy density and pressure of the null fluid, respectively. These quantities are given by:

\begin{eqnarray} \label{eq:q}
\sigma(v,r)&=&\frac{2\dot{M}(v)}{r^2},\nonumber \\
\rho(v,r)&=&\frac{2M'(v,r)}{r^2},\nonumber \\
P(v,r)&=&-\frac{M''(v,r)}{r}.
\end{eqnarray}

The Einstein field equations, combined with the equation of state \eqref{eq:defk}, yield a mass function of the form:

\begin{equation} \label{eq:solution3}
M(r,v)=M_0(v)-\frac{w_0(v)}{2}\left(Rr^2+R^2r+\frac{R^3}{2}\right)e^{-\frac{2r}{R}}.
\end{equation}

\begin{widetext}
The physical quantities \eqref{eq:q} are given by
\begin{eqnarray} \label{reg_den2}
\sigma&=&\frac{2}{r^2}\left[\dot{M}_0(v)-\frac{\dot{w}_0(v)}{2}\left(Rr^2+R^2r+\frac{R^3}{2}\right)e^{-\frac{2r}{R}}\right],\nonumber \\
\rho&=&2w_0(v)e^{-\frac{2r}{R}},\nonumber \\
P&=&2w_0(v)\left(\frac{r}{R}-1\right)e^{-\frac{2r}{R}}.
\end{eqnarray}
\end{widetext}

\subsection{Naked singularity formation}

First, we consider the case of singularity formation, where the condition

\begin{equation} \label{regular3}
w_0(v) = \frac{4M_0(v)}{R^3}
\end{equation}

is not satisfied.

A naked singularity may result from gravitational collapse if the following conditions are satisfied:

\begin{itemize}
    \item The time of singularity formation is less than the time of apparent horizon formation.
    \item There exists a family of non-spacelike, future-directed geodesics that terminate at the central singularity in the past.
\end{itemize}

The central singularity forms at \(r = 0\) at time \(v = 0\). To prove the existence of a family of non-spacelike, future-directed geodesics terminating at the central singularity in the past, we analyze the radial null geodesic, which is given by:

\begin{equation} \label{eq:radial}
\frac{dv}{dr}=\frac{2}{1-\frac{2M_0}{r}+\frac{w_0}{r}\left(r^2R+rR^2+\frac{R^3}{2}\right)e^{-\frac{2r}{R}}}.
\end{equation}

This geodesic terminates at the central singularity in the past if \(\lim\limits_{v \to 0, r \to 0} \frac{dv}{dr}\) is finite and positive. Let us denote:

\begin{equation} \label{eq:limit} 
\lim\limits_{v\rightarrow 0, r\rightarrow 0}\frac{dv}{dr}=X_0.
\end{equation}

At the time of singularity formation, i.e., at \(v = 0\), the condition \(M(0,0) = 0\) must be satisfied~\cite{Mkenyeleye:2014dwa}. Specifically, this condition implies:

\begin{equation}
M_0(0)=w_0(0)=0.
\end{equation}

If this condition is not satisfied, but we still have \(M(0,0) = 0\), it implies that \(w_0 = \frac{4M_0}{R^3}\), corresponding to the formation of a regular black hole. This scenario will be discussed in detail in the next subsection.

Thus, one can write:
\begin{eqnarray} \label{eq:sub}
M_0(v)&\sim &M_{00}v+\mathcal{O}(v^2),\nonumber \\
w_0(v)&\sim &w_{00}v+\mathcal{O}(v^2),\nonumber \\
M_{00}&\equiv &\dot{M}_0(0),\nonumber \\
w_{00}&\equiv &\dot{w}_0(0).
\end{eqnarray}

Substituting \eqref{eq:sub} into \eqref{eq:radial} and utilizing the definition \eqref{eq:limit}, we arrive at the following algebraic equation:

\begin{eqnarray} \label{eq:naked}
\xi X_0^2&-&X_0+2=0,\nonumber \\
\xi &\equiv &2M_{00}-\frac{1}{2}w_{00}R^3>0.
\end{eqnarray}

The last inequality arises from the weak energy condition, which requires \(\mu > 0\) (see \eqref{eq:q}). If \eqref{eq:naked} admits a positive root, the result of the gravitational collapse may lead to a naked singularity. Solving this quadratic equation yields:

\begin{eqnarray}
X_0^{\pm}&=&\frac{1}{2\xi}\left(1\pm \sqrt{1-8\xi}\right).
\end{eqnarray}

It can be observed that if \(\xi \leq \frac{1}{8}\), the result of gravitational collapse may lead to a naked singularity. Notably, in the Vaidya spacetime, a naked singularity forms if \(M_{00} \leq \frac{1}{16}\). However, in our model, \(M_{00}\) can exceed \(\frac{1}{16}\), leading to the following restriction:

\begin{equation}
\xi\leq \frac{1}{8}\rightarrow M_{00}\leq \frac{1}{16}+\frac{w_{00}R^3}{2}.
\end{equation}

\subsection{Regular black hole formation}

Now, we consider the formation of a regular black hole. In this scenario, the following condition must be satisfied:

\begin{equation} \label{eq:con_reg}
w_0(v) = \frac{4M_0(v)}{R^3}.
\end{equation}

A black hole is said to have a regular center if the scalar invariants remain finite in the limit \(r \to 0\). The Kretschmann scalar at the center takes the following form:
\begin{equation}
K = R_{\mu\nu\rho\sigma}R^{\mu\nu\rho\sigma} \big|_{r \to 0},
\end{equation}
where \(K\) must remain finite to ensure the regularity of the spacetime at the center.

\begin{equation}
K=\frac{32}{3}w_0^2(v)=\frac{512M_0(v)^2}{3R^6}.
\end{equation}

It is important to note that the energy flux \(\mu\) is zero at the center of the black hole, as the condition \eqref{eq:con_reg} implies \(\dot{M}(v, 0) = 0\). From an observational perspective, it is crucial to extract information from the region containing the de Sitter core. To achieve this, the following conditions must be satisfied:

\begin{itemize}
    \item The apparent horizon must be absent at \(v = 0\).
    \item There must exist a family of non-spacelike, future-directed geodesics emanating from the center of the star.
\end{itemize}

The absence of the apparent horizon as \(v \to 0\) implies that \(\lim\limits_{v \to 0} M(v, r) = 0\). This condition imposes the following relationship on the functions \(M_0(v)\) and \(w_0(v)\), i.e.,

\begin{equation}
M_0(0)=0,~~ w_0(0)=0.
\end{equation}

Now, we need to prove that as \(v \to 0\), there exists a future-directed radial null geodesic. The equation for a radial null geodesic takes the form:
\begin{equation}
\frac{dv}{dr}=\frac{2}{1-\frac{2M(v,r)}{r}}.
\end{equation}

Now, if we take the limit,
\begin{equation}
\lim\limits_{v\rightarrow 0, r\rightarrow 0} \frac{dv}{dr}\equiv X_0=2,
\end{equation}

i.e., this geodesic exists and is future-directed. Consequently, in this model, under certain physical conditions, the de Sitter core can be observed by a distant observer. 

Here, we highlight a notable property of the obtained solution, namely the phenomenon of singularity-regularity oscillation. To investigate this, let us define the function:

\begin{equation} \label{eq:defn}
N(v) \equiv w_0(v) - \frac{4M_0(v)}{R^3}.
\end{equation}

If \(N(v) \equiv 0\), the black hole remains regular. However, if \(N(v) = 0\) only at specific points \(v_1, v_2, v_3, \ldots, v_n\), the black hole evolves as follows: at \(v = 0\), a regular black hole forms. During the interval \(v \in (0, v_1)\), a singularity appears. At \(v = v_1\), the singularity vanishes, and the center becomes regular again. This cycle repeats, with a singularity appearing in intervals \(v \in (v_1, v_2)\), \(v \in (v_2, v_3)\), and so on.

As an example, let us consider a specific choice of the functions \(M_0\) and \(w_0\).
\begin{eqnarray} \label{eq:mwdef}
M_0(v)&=& \mu v,~~ \mu >0,\nonumber \\
w_0(v)&=& \nu v^2,~~ \nu >0.
\end{eqnarray}

In this case, the function \(N(v)\) equals zero at two points, \(v_1 = 0\) and \(v_2 = \frac{4\mu}{\nu R^3}\). At the time \(v = v_1\), a regular black hole forms. During the interval \(v \in (v_1, v_2)\), a singularity develops. However, at \(v = v_2\), the center becomes regular again, and for \(v > v_2\), a singular black hole forms once more.

\section{Conclusion}

In this paper, we have explored three new solutions of the Einstein field equations, namely \eqref{eq:solution1}, \eqref{eq:solution2}, and \eqref{eq:solution3}, each describing a black hole. 

The first solution, \eqref{eq:solution1}, represents a simple model of a black hole with a varying equation of state, \(P = k(r)\rho\). Depending on the parameters $M_0$ and $w_0$ , this solution can describe either a singular or a regular black hole. 

The second solution involves a black hole with a Hagedorn fluid, which is a suitable equation of state for the late stages of black hole formation. While this solution is generally difficult to express in terms of elementary functions, we analyzed a specific case, \eqref{eq:solution2}, and found it to be similar to \eqref{eq:solution1}, with the ability to describe both regular and singular black holes. Next, we examined the shadow cast by the new black hole solutions. Constraints on the black hole shadow radius (\(R_{\text{sh}}\)) for various theoretical models were derived using observations of Sagittarius A* (Sgr A*) from the Event Horizon Telescope (EHT). These constraints are summarized as follows:

\begin{itemize}
   
 \item Black Hole with De Sitter Core:
 \\  \\
    - In Fig. \ref{fig:shadow1}, the shadow radius \(r_{sh}\) is computed for varying values of \(R\). The analysis indicates that \(r_{sh}\) decreases as \(R\) increases.
     \\  \\
    - The EHT observational results for Sgr A* impose upper limits on \(R\). At a \(68\%\) confidence level (C.L.) \cite{Vagnozzi:2022moj}, the upper limit for \(R\) is determined to be \(0.9\), following the averaging of Keck and VLTI mass-to-distance ratio priors.

 \item Black Hole with Hagedorn Fluid:
  \\ \\
    - As depicted in Fig. \ref{fig:shadow2}, the shadow radius \(r_{sh}\) is calculated for varying values of \(\beta\). Initially, \(r_{sh}\) decreases with increasing \(\beta\), but it begins to increase for higher values of \(\beta\).
     \\  \\
    - Based on the EHT observations for Sgr A*, the upper limits on \(\beta\) are established. At a \(68\%\) confidence level (C.L.) \cite{Vagnozzi:2022moj}, the upper limit for \(\beta\) is found to be \(1.8\), derived from the same observational data and priors used in the De Sitter core model.

\end{itemize}
These constraints provide significant insights into the properties of black holes with different core structures and surrounding matter distributions. The horizon-scale imaging data from the EHT, particularly for Sgr A*, play a crucial role in refining these models and enhancing our understanding of black hole physics.

The first two models describe static, spherically symmetric black holes. However, real astrophysical objects gain or lose mass during processes such as gravitational collapse, accretion, or radiation. This implies that the spacetime describing real astrophysical black holes should be dynamic. For this reason, we considered the dynamical version of the solution \eqref{eq:solution1} and obtained the solution \eqref{eq:solution3}. Subsequently, we analyzed the process of gravitational collapse and discovered that it can result in the formation of a naked singularity. If we match functions $M_0(v)$ and $w_0(v)$ in such a way that a regular black hole forms then the de Sitter center can be observed by a distant observer.
The key point of the solution \eqref{eq:solution3} is that it can describe regular black hole then during the gravitational collapse it becomes singular and after some time it becomes regular again. 
The obtained solutions have an important astrophysical application:
\begin{itemize}
\item The obtained solutions can describe both regular and singular black holes, depending solely on the initial profile. Consequently, by calculating the black hole shadow, it is possible to distinguish between a singular and a regular black hole through their shadow properties. It is important to emphasize that we focus on one specific model to identify differences in the shadow characteristics.
\item 
Furthermore, by investigating the motion of $S$-stars and other objects in close proximity to the black hole, we can determine whether the central object is a singular or regular black hole.
\end{itemize}

Investigating the thermodynamical properties of the obtained solutions is particularly interesting. In particular, it is essential to examine what occurs at the moment $v = v_{\text{null}}$ and determine whether any distinctive features arise in the two regimes where the weak energy condition is either satisfied or violated. Additionally, understanding the shadow properties during the dynamical process of singular black hole formation and evaporation is crucial for gaining deeper insights into the nature of black holes. To achieve this, one needs to develop a method for analytically calculating the shadow of a dynamical black hole, a technique that is currently lacking. However, initial progress in this direction has been made in the work~\cite{Vertogradov:2024dpa}. 
The solution \eqref{eq:solution3} exhibits an unusual behavior in the Kretschmann scalar, which can be both finite and infinite at the center of the black hole during its evolution. The formation of a singularity is understandable, as many regular black hole solutions supported by non-linear electrodynamics become singular at the center during the neutralization process. However, the reverse process is counterintuitive—where a singularity transitions to a regular state during evolution—which appears unnatural. This phenomenon has not been encountered in the literature before. It requires careful investigation to either provide a physical explanation for the process or to dismiss it based on physically relevant reasoning. We leave this problem for future research.

\acknowledgments
A. {\"O}. would like to acknowledge the contribution of the COST Action CA21106 - COSMIC WISPers in the Dark Universe: Theory, astrophysics and experiments (CosmicWISPers) and the COST Action CA22113 - Fundamental challenges in theoretical physics (THEORY-CHALLENGES). We also thank TUBITAK and SCOAP3 for their support.

\bibliographystyle{apsrev4-1}
\bibliography{ref}

\begin{thebibliography}{84}%
\makeatletter
\providecommand \@ifxundefined [1]{%
 \@ifx{#1\undefined}
}%
\providecommand \@ifnum [1]{%
 \ifnum #1\expandafter \@firstoftwo
 \else \expandafter \@secondoftwo
 \fi
}%
\providecommand \@ifx [1]{%
 \ifx #1\expandafter \@firstoftwo
 \else \expandafter \@secondoftwo
 \fi
}%
\providecommand \natexlab [1]{#1}%
\providecommand \enquote  [1]{``#1''}%
\providecommand \bibnamefont  [1]{#1}%
\providecommand \bibfnamefont [1]{#1}%
\providecommand \citenamefont [1]{#1}%
\providecommand \href@noop [0]{\@secondoftwo}%
\providecommand \href [0]{\begingroup \@sanitize@url \@href}%
\providecommand \@href[1]{\@@startlink{#1}\@@href}%
\providecommand \@@href[1]{\endgroup#1\@@endlink}%
\providecommand \@sanitize@url [0]{\catcode `\\12\catcode `\$12\catcode `\&12\catcode `\#12\catcode `\^12\catcode `\_12\catcode `\%12\relax}%
\providecommand \@@startlink[1]{}%
\providecommand \@@endlink[0]{}%
\providecommand \url  [0]{\begingroup\@sanitize@url \@url }%
\providecommand \@url [1]{\endgroup\@href {#1}{\urlprefix }}%
\providecommand \urlprefix  [0]{URL }%
\providecommand \Eprint [0]{\href }%
\providecommand \doibase [0]{http://dx.doi.org/}%
\providecommand \selectlanguage [0]{\@gobble}%
\providecommand \bibinfo  [0]{\@secondoftwo}%
\providecommand \bibfield  [0]{\@secondoftwo}%
\providecommand \translation [1]{[#1]}%
\providecommand \BibitemOpen [0]{}%
\providecommand \bibitemStop [0]{}%
\providecommand \bibitemNoStop [0]{.\EOS\space}%
\providecommand \EOS [0]{\spacefactor3000\relax}%
\providecommand \BibitemShut  [1]{\csname bibitem#1\endcsname}%
\let\auto@bib@innerbib\@empty
\bibitem [{\citenamefont {Oppenheimer}\ and\ \citenamefont {Snyder}(1939)}]{Oppenheimer:1939ue}%
  \BibitemOpen
  \bibfield  {author} {\bibinfo {author} {\bibfnamefont {J.~R.}\ \bibnamefont {Oppenheimer}}\ and\ \bibinfo {author} {\bibfnamefont {H.}~\bibnamefont {Snyder}},\ }\href {\doibase 10.1103/PhysRev.56.455} {\bibfield  {journal} {\bibinfo  {journal} {Phys. Rev.}\ }\textbf {\bibinfo {volume} {56}},\ \bibinfo {pages} {455} (\bibinfo {year} {1939})}\BibitemShut {NoStop}%
\bibitem [{\citenamefont {{Datt}}(1938)}]{datt}%
  \BibitemOpen
  \bibfield  {author} {\bibinfo {author} {\bibfnamefont {B.}~\bibnamefont {{Datt}}},\ }\href {\doibase 10.1007/BF01374951} {\bibfield  {journal} {\bibinfo  {journal} {Zeitschrift fur Physik}\ }\textbf {\bibinfo {volume} {108}},\ \bibinfo {pages} {314} (\bibinfo {year} {1938})}\BibitemShut {NoStop}%
\bibitem [{\citenamefont {Penrose}(1969)}]{Penrose:1969pc}%
  \BibitemOpen
  \bibfield  {author} {\bibinfo {author} {\bibfnamefont {R.}~\bibnamefont {Penrose}},\ }\href {\doibase 10.1023/A:1016578408204} {\bibfield  {journal} {\bibinfo  {journal} {Riv. Nuovo Cim.}\ }\textbf {\bibinfo {volume} {1}},\ \bibinfo {pages} {252} (\bibinfo {year} {1969})}\BibitemShut {NoStop}%
\bibitem [{\citenamefont {Misyura}\ \emph {et~al.}(2024)\citenamefont {Misyura}, \citenamefont {Rincon},\ and\ \citenamefont {Vertogradov}}]{Misyura:2024fho}%
  \BibitemOpen
  \bibfield  {author} {\bibinfo {author} {\bibfnamefont {M.}~\bibnamefont {Misyura}}, \bibinfo {author} {\bibfnamefont {A.}~\bibnamefont {Rincon}}, \ and\ \bibinfo {author} {\bibfnamefont {V.}~\bibnamefont {Vertogradov}},\ }\href@noop {} {\  (\bibinfo {year} {2024})},\ \Eprint {http://arxiv.org/abs/2405.05370} {arXiv:2405.05370 [gr-qc]} \BibitemShut {NoStop}%
\bibitem [{\citenamefont {Konoplich}\ \emph {et~al.}(1999)\citenamefont {Konoplich}, \citenamefont {Rubin}, \citenamefont {Sakharov},\ and\ \citenamefont {Khlopov}}]{Konoplich:1999qq}%
  \BibitemOpen
  \bibfield  {author} {\bibinfo {author} {\bibfnamefont {R.~V.}\ \bibnamefont {Konoplich}}, \bibinfo {author} {\bibfnamefont {S.~G.}\ \bibnamefont {Rubin}}, \bibinfo {author} {\bibfnamefont {A.~S.}\ \bibnamefont {Sakharov}}, \ and\ \bibinfo {author} {\bibfnamefont {M.~Y.}\ \bibnamefont {Khlopov}},\ }\href@noop {} {\bibfield  {journal} {\bibinfo  {journal} {Phys. Atom. Nucl.}\ }\textbf {\bibinfo {volume} {62}},\ \bibinfo {pages} {1593} (\bibinfo {year} {1999})}\BibitemShut {NoStop}%
\bibitem [{\citenamefont {Khlopov}\ \emph {et~al.}(1999)\citenamefont {Khlopov}, \citenamefont {Konoplich}, \citenamefont {Rubin},\ and\ \citenamefont {Sakharov}}]{Khlopov:1999ys}%
  \BibitemOpen
  \bibfield  {author} {\bibinfo {author} {\bibfnamefont {M.~Y.}\ \bibnamefont {Khlopov}}, \bibinfo {author} {\bibfnamefont {R.~V.}\ \bibnamefont {Konoplich}}, \bibinfo {author} {\bibfnamefont {S.~G.}\ \bibnamefont {Rubin}}, \ and\ \bibinfo {author} {\bibfnamefont {A.~S.}\ \bibnamefont {Sakharov}},\ }\href@noop {} {\bibfield  {journal} {\bibinfo  {journal} {Grav. Cosmol.}\ }\textbf {\bibinfo {volume} {2}},\ \bibinfo {pages} {S1} (\bibinfo {year} {1999})},\ \Eprint {http://arxiv.org/abs/hep-ph/9912422} {arXiv:hep-ph/9912422} \BibitemShut {NoStop}%
\bibitem [{\citenamefont {Khlopov}(2010)}]{Khlopov:2008qy}%
  \BibitemOpen
  \bibfield  {author} {\bibinfo {author} {\bibfnamefont {M.~Y.}\ \bibnamefont {Khlopov}},\ }\href {\doibase 10.1088/1674-4527/10/6/001} {\bibfield  {journal} {\bibinfo  {journal} {Res. Astron. Astrophys.}\ }\textbf {\bibinfo {volume} {10}},\ \bibinfo {pages} {495} (\bibinfo {year} {2010})},\ \Eprint {http://arxiv.org/abs/0801.0116} {arXiv:0801.0116 [astro-ph]} \BibitemShut {NoStop}%
\bibitem [{\citenamefont {Belotsky}\ \emph {et~al.}(2014)\citenamefont {Belotsky}, \citenamefont {Dmitriev}, \citenamefont {Esipova}, \citenamefont {Gani}, \citenamefont {Grobov}, \citenamefont {Khlopov}, \citenamefont {Kirillov}, \citenamefont {Rubin},\ and\ \citenamefont {Svadkovsky}}]{Belotsky:2014kca}%
  \BibitemOpen
  \bibfield  {author} {\bibinfo {author} {\bibfnamefont {K.~M.}\ \bibnamefont {Belotsky}}, \bibinfo {author} {\bibfnamefont {A.~D.}\ \bibnamefont {Dmitriev}}, \bibinfo {author} {\bibfnamefont {E.~A.}\ \bibnamefont {Esipova}}, \bibinfo {author} {\bibfnamefont {V.~A.}\ \bibnamefont {Gani}}, \bibinfo {author} {\bibfnamefont {A.~V.}\ \bibnamefont {Grobov}}, \bibinfo {author} {\bibfnamefont {M.~Y.}\ \bibnamefont {Khlopov}}, \bibinfo {author} {\bibfnamefont {A.~A.}\ \bibnamefont {Kirillov}}, \bibinfo {author} {\bibfnamefont {S.~G.}\ \bibnamefont {Rubin}}, \ and\ \bibinfo {author} {\bibfnamefont {I.~V.}\ \bibnamefont {Svadkovsky}},\ }\href {\doibase 10.1142/S0217732314400057} {\bibfield  {journal} {\bibinfo  {journal} {Mod. Phys. Lett. A}\ }\textbf {\bibinfo {volume} {29}},\ \bibinfo {pages} {1440005} (\bibinfo {year} {2014})},\ \Eprint {http://arxiv.org/abs/1410.0203} {arXiv:1410.0203 [astro-ph.CO]} \BibitemShut {NoStop}%
\bibitem [{\citenamefont {Dymnikova}\ and\ \citenamefont {Khlopov}(2015)}]{Dymnikova:2015yma}%
  \BibitemOpen
  \bibfield  {author} {\bibinfo {author} {\bibfnamefont {I.}~\bibnamefont {Dymnikova}}\ and\ \bibinfo {author} {\bibfnamefont {M.}~\bibnamefont {Khlopov}},\ }\href {\doibase 10.1142/S0218271815450029} {\bibfield  {journal} {\bibinfo  {journal} {Int. J. Mod. Phys. D}\ }\textbf {\bibinfo {volume} {24}},\ \bibinfo {pages} {1545002} (\bibinfo {year} {2015})},\ \Eprint {http://arxiv.org/abs/1510.01351} {arXiv:1510.01351 [gr-qc]} \BibitemShut {NoStop}%
\bibitem [{\citenamefont {Stephani}\ \emph {et~al.}(2003)\citenamefont {Stephani}, \citenamefont {Kramer}, \citenamefont {MacCallum}, \citenamefont {Hoenselaers},\ and\ \citenamefont {Herlt}}]{Stephani:2003tm}%
  \BibitemOpen
  \bibfield  {author} {\bibinfo {author} {\bibfnamefont {H.}~\bibnamefont {Stephani}}, \bibinfo {author} {\bibfnamefont {D.}~\bibnamefont {Kramer}}, \bibinfo {author} {\bibfnamefont {M.~A.~H.}\ \bibnamefont {MacCallum}}, \bibinfo {author} {\bibfnamefont {C.}~\bibnamefont {Hoenselaers}}, \ and\ \bibinfo {author} {\bibfnamefont {E.}~\bibnamefont {Herlt}},\ }\href {\doibase 10.1017/CBO9780511535185} {\emph {\bibinfo {title} {{Exact solutions of Einstein's field equations}}}},\ Cambridge Monographs on Mathematical Physics\ (\bibinfo  {publisher} {Cambridge Univ. Press},\ \bibinfo {address} {Cambridge},\ \bibinfo {year} {2003})\BibitemShut {NoStop}%
\bibitem [{\citenamefont {Delgaty}\ and\ \citenamefont {Lake}(1998)}]{Delgaty:1998uy}%
  \BibitemOpen
  \bibfield  {author} {\bibinfo {author} {\bibfnamefont {M.~S.~R.}\ \bibnamefont {Delgaty}}\ and\ \bibinfo {author} {\bibfnamefont {K.}~\bibnamefont {Lake}},\ }\href {\doibase 10.1016/S0010-4655(98)00130-1} {\bibfield  {journal} {\bibinfo  {journal} {Comput. Phys. Commun.}\ }\textbf {\bibinfo {volume} {115}},\ \bibinfo {pages} {395} (\bibinfo {year} {1998})},\ \Eprint {http://arxiv.org/abs/gr-qc/9809013} {arXiv:gr-qc/9809013} \BibitemShut {NoStop}%
\bibitem [{\citenamefont {Semiz}(2011)}]{Semiz:2008ny}%
  \BibitemOpen
  \bibfield  {author} {\bibinfo {author} {\bibfnamefont {I.}~\bibnamefont {Semiz}},\ }\href {\doibase 10.1142/S0129055X1100445X} {\bibfield  {journal} {\bibinfo  {journal} {Rev. Math. Phys.}\ }\textbf {\bibinfo {volume} {23}},\ \bibinfo {pages} {865} (\bibinfo {year} {2011})},\ \Eprint {http://arxiv.org/abs/0810.0634} {arXiv:0810.0634 [gr-qc]} \BibitemShut {NoStop}%
\bibitem [{\citenamefont {Herrera}\ and\ \citenamefont {Santos}(1997)}]{Herrera:1997plx}%
  \BibitemOpen
  \bibfield  {author} {\bibinfo {author} {\bibfnamefont {L.}~\bibnamefont {Herrera}}\ and\ \bibinfo {author} {\bibfnamefont {N.~O.}\ \bibnamefont {Santos}},\ }\href {\doibase 10.1016/S0370-1573(96)00042-7} {\bibfield  {journal} {\bibinfo  {journal} {Phys. Rept.}\ }\textbf {\bibinfo {volume} {286}},\ \bibinfo {pages} {53} (\bibinfo {year} {1997})}\BibitemShut {NoStop}%
\bibitem [{\citenamefont {Ruderman}(1972)}]{Ruderman:1972aj}%
  \BibitemOpen
  \bibfield  {author} {\bibinfo {author} {\bibfnamefont {M.}~\bibnamefont {Ruderman}},\ }\href {\doibase 10.1146/annurev.aa.10.090172.002235} {\bibfield  {journal} {\bibinfo  {journal} {Ann. Rev. Astron. Astrophys.}\ }\textbf {\bibinfo {volume} {10}},\ \bibinfo {pages} {427} (\bibinfo {year} {1972})}\BibitemShut {NoStop}%
\bibitem [{\citenamefont {Kim}(2017)}]{Kim:2017hem}%
  \BibitemOpen
  \bibfield  {author} {\bibinfo {author} {\bibfnamefont {H.-C.}\ \bibnamefont {Kim}},\ }\href {\doibase 10.1103/PhysRevD.96.064053} {\bibfield  {journal} {\bibinfo  {journal} {Phys. Rev. D}\ }\textbf {\bibinfo {volume} {96}},\ \bibinfo {pages} {064053} (\bibinfo {year} {2017})},\ \Eprint {http://arxiv.org/abs/1708.02373} {arXiv:1708.02373 [gr-qc]} \BibitemShut {NoStop}%
\bibitem [{\citenamefont {{Sakharov}}(1966)}]{Sakharov1966JETP...22..241S}%
  \BibitemOpen
  \bibfield  {author} {\bibinfo {author} {\bibfnamefont {A.~D.}\ \bibnamefont {{Sakharov}}},\ }\href@noop {} {\bibfield  {journal} {\bibinfo  {journal} {Soviet Journal of Experimental and Theoretical Physics}\ }\textbf {\bibinfo {volume} {22}},\ \bibinfo {pages} {241} (\bibinfo {year} {1966})}\BibitemShut {NoStop}%
\bibitem [{\citenamefont {{Gliner}}(1966)}]{Gliner1966JETP...22..378G}%
  \BibitemOpen
  \bibfield  {author} {\bibinfo {author} {\bibfnamefont {E.~B.}\ \bibnamefont {{Gliner}}},\ }\href@noop {} {\bibfield  {journal} {\bibinfo  {journal} {Soviet Journal of Experimental and Theoretical Physics}\ }\textbf {\bibinfo {volume} {22}},\ \bibinfo {pages} {378} (\bibinfo {year} {1966})}\BibitemShut {NoStop}%
\bibitem [{\citenamefont {Dymnikova}(1992)}]{Dymnikova:1992ux}%
  \BibitemOpen
  \bibfield  {author} {\bibinfo {author} {\bibfnamefont {I.}~\bibnamefont {Dymnikova}},\ }\href {\doibase 10.1007/BF00760226} {\bibfield  {journal} {\bibinfo  {journal} {Gen. Rel. Grav.}\ }\textbf {\bibinfo {volume} {24}},\ \bibinfo {pages} {235} (\bibinfo {year} {1992})}\BibitemShut {NoStop}%
\bibitem [{\citenamefont {{Gurevich}}(1975)}]{Gurevich1975Ap&SS..38...67G}%
  \BibitemOpen
  \bibfield  {author} {\bibinfo {author} {\bibfnamefont {L.~E.}\ \bibnamefont {{Gurevich}}},\ }\href {\doibase 10.1007/BF00646099} {\bibfield  {journal} {\bibinfo  {journal} {Astrophys Space Sci}\ }\textbf {\bibinfo {volume} {38}},\ \bibinfo {pages} {67–78} (\bibinfo {year} {1975})}\BibitemShut {NoStop}%
\bibitem [{\citenamefont {Starobinsky}(1979)}]{Starobinsky:1979ty}%
  \BibitemOpen
  \bibfield  {author} {\bibinfo {author} {\bibfnamefont {A.~A.}\ \bibnamefont {Starobinsky}},\ }\href@noop {} {\bibfield  {journal} {\bibinfo  {journal} {JETP Lett.}\ }\textbf {\bibinfo {volume} {30}},\ \bibinfo {pages} {682} (\bibinfo {year} {1979})}\BibitemShut {NoStop}%
\bibitem [{\citenamefont {{Bardeen}}(1968)}]{Bardeen1968qtr..conf...87B}%
  \BibitemOpen
  \bibfield  {author} {\bibinfo {author} {\bibfnamefont {J.}~\bibnamefont {{Bardeen}}},\ }in\ \href@noop {} {\emph {\bibinfo {booktitle} {Proceedings of the 5th International Conference on Gravitation and the Theory of Relativity}}}\ (\bibinfo {year} {1968})\ p.~\bibinfo {pages} {87}\BibitemShut {NoStop}%
\bibitem [{\citenamefont {Ansoldi}(2008)}]{Ansoldi:2008jw}%
  \BibitemOpen
  \bibfield  {author} {\bibinfo {author} {\bibfnamefont {S.}~\bibnamefont {Ansoldi}},\ }in\ \href@noop {} {\emph {\bibinfo {booktitle} {{Conference on Black Holes and Naked Singularities}}}}\ (\bibinfo {year} {2008})\ \Eprint {http://arxiv.org/abs/0802.0330} {arXiv:0802.0330 [gr-qc]} \BibitemShut {NoStop}%
\bibitem [{\citenamefont {Bambi}(2023)}]{Bambi:2023try}%
  \BibitemOpen
  \bibinfo {editor} {\bibfnamefont {C.}~\bibnamefont {Bambi}},\ ed.,\ \href {\doibase 10.1007/978-981-99-1596-5} {\emph {\bibinfo {title} {{Regular Black Holes. Towards a New Paradigm of Gravitational Collapse}}}},\ Springer Series in Astrophysics and Cosmology\ (\bibinfo  {publisher} {Springer},\ \bibinfo {year} {2023})\ \Eprint {http://arxiv.org/abs/2307.13249} {arXiv:2307.13249 [gr-qc]} \BibitemShut {NoStop}%
\bibitem [{\citenamefont {Lan}\ \emph {et~al.}(2023)\citenamefont {Lan}, \citenamefont {Yang}, \citenamefont {Guo},\ and\ \citenamefont {Miao}}]{Lan:2023cvz}%
  \BibitemOpen
  \bibfield  {author} {\bibinfo {author} {\bibfnamefont {C.}~\bibnamefont {Lan}}, \bibinfo {author} {\bibfnamefont {H.}~\bibnamefont {Yang}}, \bibinfo {author} {\bibfnamefont {Y.}~\bibnamefont {Guo}}, \ and\ \bibinfo {author} {\bibfnamefont {Y.-G.}\ \bibnamefont {Miao}},\ }\href {\doibase 10.1007/s10773-023-05454-1} {\bibfield  {journal} {\bibinfo  {journal} {Int. J. Theor. Phys.}\ }\textbf {\bibinfo {volume} {62}},\ \bibinfo {pages} {202} (\bibinfo {year} {2023})},\ \Eprint {http://arxiv.org/abs/2303.11696} {arXiv:2303.11696 [gr-qc]} \BibitemShut {NoStop}%
\bibitem [{\citenamefont {Ayon-Beato}\ and\ \citenamefont {Garcia}(1998)}]{Ayon-Beato:1998hmi}%
  \BibitemOpen
  \bibfield  {author} {\bibinfo {author} {\bibfnamefont {E.}~\bibnamefont {Ayon-Beato}}\ and\ \bibinfo {author} {\bibfnamefont {A.}~\bibnamefont {Garcia}},\ }\href {\doibase 10.1103/PhysRevLett.80.5056} {\bibfield  {journal} {\bibinfo  {journal} {Phys. Rev. Lett.}\ }\textbf {\bibinfo {volume} {80}},\ \bibinfo {pages} {5056} (\bibinfo {year} {1998})},\ \Eprint {http://arxiv.org/abs/gr-qc/9911046} {arXiv:gr-qc/9911046} \BibitemShut {NoStop}%
\bibitem [{\citenamefont {Ayon-Beato}\ and\ \citenamefont {Garcia}(1999)}]{Ayon-Beato:1999qin}%
  \BibitemOpen
  \bibfield  {author} {\bibinfo {author} {\bibfnamefont {E.}~\bibnamefont {Ayon-Beato}}\ and\ \bibinfo {author} {\bibfnamefont {A.}~\bibnamefont {Garcia}},\ }\href {\doibase 10.1023/A:1026640911319} {\bibfield  {journal} {\bibinfo  {journal} {Gen. Rel. Grav.}\ }\textbf {\bibinfo {volume} {31}},\ \bibinfo {pages} {629} (\bibinfo {year} {1999})},\ \Eprint {http://arxiv.org/abs/gr-qc/9911084} {arXiv:gr-qc/9911084} \BibitemShut {NoStop}%
\bibitem [{\citenamefont {Singh}\ \emph {et~al.}(2022{\natexlab{a}})\citenamefont {Singh}, \citenamefont {Ghosh},\ and\ \citenamefont {Maharaj}}]{Singh:2022xgi}%
  \BibitemOpen
  \bibfield  {author} {\bibinfo {author} {\bibfnamefont {D.~V.}\ \bibnamefont {Singh}}, \bibinfo {author} {\bibfnamefont {S.~G.}\ \bibnamefont {Ghosh}}, \ and\ \bibinfo {author} {\bibfnamefont {S.~D.}\ \bibnamefont {Maharaj}},\ }\href {\doibase 10.1016/j.nuclphysb.2022.115854} {\bibfield  {journal} {\bibinfo  {journal} {Nucl. Phys. B}\ }\textbf {\bibinfo {volume} {981}},\ \bibinfo {pages} {115854} (\bibinfo {year} {2022}{\natexlab{a}})}\BibitemShut {NoStop}%
\bibitem [{\citenamefont {Singh}\ \emph {et~al.}(2022{\natexlab{b}})\citenamefont {Singh}, \citenamefont {Shukla},\ and\ \citenamefont {Upadhyay}}]{Singh:2022ycn}%
  \BibitemOpen
  \bibfield  {author} {\bibinfo {author} {\bibfnamefont {D.~V.}\ \bibnamefont {Singh}}, \bibinfo {author} {\bibfnamefont {A.}~\bibnamefont {Shukla}}, \ and\ \bibinfo {author} {\bibfnamefont {S.}~\bibnamefont {Upadhyay}},\ }\href {\doibase 10.1016/j.aop.2022.169157} {\bibfield  {journal} {\bibinfo  {journal} {Annals Phys.}\ }\textbf {\bibinfo {volume} {447}},\ \bibinfo {pages} {169157} (\bibinfo {year} {2022}{\natexlab{b}})},\ \Eprint {http://arxiv.org/abs/2211.09673} {arXiv:2211.09673 [gr-qc]} \BibitemShut {NoStop}%
\bibitem [{\citenamefont {Sudhanshu}\ \emph {et~al.}(2024)\citenamefont {Sudhanshu}, \citenamefont {Singh}, \citenamefont {Upadhyay}, \citenamefont {Myrzakulov},\ and\ \citenamefont {Myrzakulov}}]{Sudhanshu:2024wqb}%
  \BibitemOpen
  \bibfield  {author} {\bibinfo {author} {\bibfnamefont {H.~K.}\ \bibnamefont {Sudhanshu}}, \bibinfo {author} {\bibfnamefont {D.~V.}\ \bibnamefont {Singh}}, \bibinfo {author} {\bibfnamefont {S.}~\bibnamefont {Upadhyay}}, \bibinfo {author} {\bibfnamefont {Y.}~\bibnamefont {Myrzakulov}}, \ and\ \bibinfo {author} {\bibfnamefont {K.}~\bibnamefont {Myrzakulov}},\ }\href {\doibase 10.1016/j.dark.2024.101648} {\bibfield  {journal} {\bibinfo  {journal} {Phys. Dark Univ.}\ }\textbf {\bibinfo {volume} {46}},\ \bibinfo {pages} {101648} (\bibinfo {year} {2024})},\ \Eprint {http://arxiv.org/abs/2410.04174} {arXiv:2410.04174 [gr-qc]} \BibitemShut {NoStop}%
\bibitem [{\citenamefont {Hagedorn}(1965)}]{Hagedorn:1965st}%
  \BibitemOpen
  \bibfield  {author} {\bibinfo {author} {\bibfnamefont {R.}~\bibnamefont {Hagedorn}},\ }\href@noop {} {\bibfield  {journal} {\bibinfo  {journal} {Nuovo Cim. Suppl.}\ }\textbf {\bibinfo {volume} {3}},\ \bibinfo {pages} {147} (\bibinfo {year} {1965})}\BibitemShut {NoStop}%
\bibitem [{\citenamefont {Atick}\ and\ \citenamefont {Witten}(1988)}]{Atick:1988si}%
  \BibitemOpen
  \bibfield  {author} {\bibinfo {author} {\bibfnamefont {J.~J.}\ \bibnamefont {Atick}}\ and\ \bibinfo {author} {\bibfnamefont {E.}~\bibnamefont {Witten}},\ }\href {\doibase 10.1016/0550-3213(88)90151-4} {\bibfield  {journal} {\bibinfo  {journal} {Nucl. Phys. B}\ }\textbf {\bibinfo {volume} {310}},\ \bibinfo {pages} {291} (\bibinfo {year} {1988})}\BibitemShut {NoStop}%
\bibitem [{\citenamefont {Giddings}(1989)}]{Giddings:1989xe}%
  \BibitemOpen
  \bibfield  {author} {\bibinfo {author} {\bibfnamefont {S.~B.}\ \bibnamefont {Giddings}},\ }\href {\doibase 10.1016/0370-2693(89)90288-8} {\bibfield  {journal} {\bibinfo  {journal} {Phys. Lett. B}\ }\textbf {\bibinfo {volume} {226}},\ \bibinfo {pages} {55} (\bibinfo {year} {1989})}\BibitemShut {NoStop}%
\bibitem [{\citenamefont {Grignani}\ \emph {et~al.}(2001)\citenamefont {Grignani}, \citenamefont {Orselli},\ and\ \citenamefont {Semenoff}}]{Grignani:2001ik}%
  \BibitemOpen
  \bibfield  {author} {\bibinfo {author} {\bibfnamefont {G.}~\bibnamefont {Grignani}}, \bibinfo {author} {\bibfnamefont {M.}~\bibnamefont {Orselli}}, \ and\ \bibinfo {author} {\bibfnamefont {G.~W.}\ \bibnamefont {Semenoff}},\ }\href {\doibase 10.1088/1126-6708/2001/11/058} {\bibfield  {journal} {\bibinfo  {journal} {JHEP}\ }\textbf {\bibinfo {volume} {11}},\ \bibinfo {pages} {058} (\bibinfo {year} {2001})},\ \Eprint {http://arxiv.org/abs/hep-th/0110152} {arXiv:hep-th/0110152} \BibitemShut {NoStop}%
\bibitem [{\citenamefont {Maggiore}(1998)}]{Maggiore:1997vw}%
  \BibitemOpen
  \bibfield  {author} {\bibinfo {author} {\bibfnamefont {M.}~\bibnamefont {Maggiore}},\ }\href {\doibase 10.1016/S0550-3213(98)00362-9} {\bibfield  {journal} {\bibinfo  {journal} {Nucl. Phys. B}\ }\textbf {\bibinfo {volume} {525}},\ \bibinfo {pages} {413} (\bibinfo {year} {1998})},\ \Eprint {http://arxiv.org/abs/gr-qc/9709004} {arXiv:gr-qc/9709004} \BibitemShut {NoStop}%
\bibitem [{\citenamefont {Magueijo}\ and\ \citenamefont {Pogosian}(2003)}]{Magueijo:2002pg}%
  \BibitemOpen
  \bibfield  {author} {\bibinfo {author} {\bibfnamefont {J.}~\bibnamefont {Magueijo}}\ and\ \bibinfo {author} {\bibfnamefont {L.}~\bibnamefont {Pogosian}},\ }\href {\doibase 10.1103/PhysRevD.67.043518} {\bibfield  {journal} {\bibinfo  {journal} {Phys. Rev. D}\ }\textbf {\bibinfo {volume} {67}},\ \bibinfo {pages} {043518} (\bibinfo {year} {2003})},\ \Eprint {http://arxiv.org/abs/astro-ph/0211337} {arXiv:astro-ph/0211337} \BibitemShut {NoStop}%
\bibitem [{\citenamefont {Bassett}\ \emph {et~al.}(2003)\citenamefont {Bassett}, \citenamefont {Borunda}, \citenamefont {Serone},\ and\ \citenamefont {Tsujikawa}}]{Bassett:2003ck}%
  \BibitemOpen
  \bibfield  {author} {\bibinfo {author} {\bibfnamefont {B.~A.}\ \bibnamefont {Bassett}}, \bibinfo {author} {\bibfnamefont {M.}~\bibnamefont {Borunda}}, \bibinfo {author} {\bibfnamefont {M.}~\bibnamefont {Serone}}, \ and\ \bibinfo {author} {\bibfnamefont {S.}~\bibnamefont {Tsujikawa}},\ }\href {\doibase 10.1103/PhysRevD.67.123506} {\bibfield  {journal} {\bibinfo  {journal} {Phys. Rev. D}\ }\textbf {\bibinfo {volume} {67}},\ \bibinfo {pages} {123506} (\bibinfo {year} {2003})},\ \Eprint {http://arxiv.org/abs/hep-th/0301180} {arXiv:hep-th/0301180} \BibitemShut {NoStop}%
\bibitem [{\citenamefont {Abbott}\ \emph {et~al.}(2016)\citenamefont {Abbott} \emph {et~al.}}]{LIGOScientific:2016aoc}%
  \BibitemOpen
  \bibfield  {author} {\bibinfo {author} {\bibfnamefont {B.~P.}\ \bibnamefont {Abbott}} \emph {et~al.} (\bibinfo {collaboration} {LIGO Scientific, Virgo}),\ }\href {\doibase 10.1103/PhysRevLett.116.061102} {\bibfield  {journal} {\bibinfo  {journal} {Phys. Rev. Lett.}\ }\textbf {\bibinfo {volume} {116}},\ \bibinfo {pages} {061102} (\bibinfo {year} {2016})},\ \Eprint {http://arxiv.org/abs/1602.03837} {arXiv:1602.03837 [gr-qc]} \BibitemShut {NoStop}%
\bibitem [{\citenamefont {Akiyama}\ \emph {et~al.}(2019)\citenamefont {Akiyama} \emph {et~al.}}]{EventHorizonTelescope:2019dse}%
  \BibitemOpen
  \bibfield  {author} {\bibinfo {author} {\bibfnamefont {K.}~\bibnamefont {Akiyama}} \emph {et~al.} (\bibinfo {collaboration} {Event Horizon Telescope}),\ }\href {\doibase 10.3847/2041-8213/ab0ec7} {\bibfield  {journal} {\bibinfo  {journal} {Astrophys. J. Lett.}\ }\textbf {\bibinfo {volume} {875}},\ \bibinfo {pages} {L1} (\bibinfo {year} {2019})},\ \Eprint {http://arxiv.org/abs/1906.11238} {arXiv:1906.11238 [astro-ph.GA]} \BibitemShut {NoStop}%
\bibitem [{\citenamefont {Akiyama}\ \emph {et~al.}(2022)\citenamefont {Akiyama} \emph {et~al.}}]{EventHorizonTelescope:2022wkp}%
  \BibitemOpen
  \bibfield  {author} {\bibinfo {author} {\bibfnamefont {K.}~\bibnamefont {Akiyama}} \emph {et~al.} (\bibinfo {collaboration} {Event Horizon Telescope}),\ }\href {\doibase 10.3847/2041-8213/ac6674} {\bibfield  {journal} {\bibinfo  {journal} {Astrophys. J. Lett.}\ }\textbf {\bibinfo {volume} {930}},\ \bibinfo {pages} {L12} (\bibinfo {year} {2022})},\ \Eprint {http://arxiv.org/abs/2311.08680} {arXiv:2311.08680 [astro-ph.HE]} \BibitemShut {NoStop}%
\bibitem [{\citenamefont {Pantig}\ \emph {et~al.}(2022)\citenamefont {Pantig}, \citenamefont {Mastrototaro}, \citenamefont {Lambiase},\ and\ \citenamefont {\"Ovg\"un}}]{Pantig:2022gih}%
  \BibitemOpen
  \bibfield  {author} {\bibinfo {author} {\bibfnamefont {R.~C.}\ \bibnamefont {Pantig}}, \bibinfo {author} {\bibfnamefont {L.}~\bibnamefont {Mastrototaro}}, \bibinfo {author} {\bibfnamefont {G.}~\bibnamefont {Lambiase}}, \ and\ \bibinfo {author} {\bibfnamefont {A.}~\bibnamefont {\"Ovg\"un}},\ }\href {\doibase 10.1140/epjc/s10052-022-11125-y} {\bibfield  {journal} {\bibinfo  {journal} {Eur. Phys. J. C}\ }\textbf {\bibinfo {volume} {82}},\ \bibinfo {pages} {1155} (\bibinfo {year} {2022})},\ \Eprint {http://arxiv.org/abs/2208.06664} {arXiv:2208.06664 [gr-qc]} \BibitemShut {NoStop}%
\bibitem [{\citenamefont {Pantig}\ and\ \citenamefont {\"Ovg\"un}(2023)}]{Pantig:2022sjb}%
  \BibitemOpen
  \bibfield  {author} {\bibinfo {author} {\bibfnamefont {R.~C.}\ \bibnamefont {Pantig}}\ and\ \bibinfo {author} {\bibfnamefont {A.}~\bibnamefont {\"Ovg\"un}},\ }\href {\doibase 10.1002/prop.202200164} {\bibfield  {journal} {\bibinfo  {journal} {Fortsch. Phys.}\ }\textbf {\bibinfo {volume} {71}},\ \bibinfo {pages} {2200164} (\bibinfo {year} {2023})},\ \Eprint {http://arxiv.org/abs/2210.00523} {arXiv:2210.00523 [gr-qc]} \BibitemShut {NoStop}%
\bibitem [{\citenamefont {Kuang}\ and\ \citenamefont {\"Ovg\"un}(2022)}]{Kuang:2022xjp}%
  \BibitemOpen
  \bibfield  {author} {\bibinfo {author} {\bibfnamefont {X.-M.}\ \bibnamefont {Kuang}}\ and\ \bibinfo {author} {\bibfnamefont {A.}~\bibnamefont {\"Ovg\"un}},\ }\href {\doibase 10.1016/j.aop.2022.169147} {\bibfield  {journal} {\bibinfo  {journal} {Annals Phys.}\ }\textbf {\bibinfo {volume} {447}},\ \bibinfo {pages} {169147} (\bibinfo {year} {2022})},\ \Eprint {http://arxiv.org/abs/2205.11003} {arXiv:2205.11003 [gr-qc]} \BibitemShut {NoStop}%
\bibitem [{\citenamefont {Zakharov}(2018)}]{Zakharov:2018awx}%
  \BibitemOpen
  \bibfield  {author} {\bibinfo {author} {\bibfnamefont {A.~F.}\ \bibnamefont {Zakharov}},\ }\href {\doibase 10.1140/epjc/s10052-018-6166-5} {\bibfield  {journal} {\bibinfo  {journal} {Eur. Phys. J. C}\ }\textbf {\bibinfo {volume} {78}},\ \bibinfo {pages} {689} (\bibinfo {year} {2018})},\ \Eprint {http://arxiv.org/abs/1804.10374} {arXiv:1804.10374 [gr-qc]} \BibitemShut {NoStop}%
\bibitem [{\citenamefont {Zakharov}\ \emph {et~al.}(2018)\citenamefont {Zakharov}, \citenamefont {Jovanovi\'c}, \citenamefont {Borka},\ and\ \citenamefont {Borka~Jovanovi\'c}}]{Zakharov:2018cbj}%
  \BibitemOpen
  \bibfield  {author} {\bibinfo {author} {\bibfnamefont {A.~F.}\ \bibnamefont {Zakharov}}, \bibinfo {author} {\bibfnamefont {P.}~\bibnamefont {Jovanovi\'c}}, \bibinfo {author} {\bibfnamefont {D.}~\bibnamefont {Borka}}, \ and\ \bibinfo {author} {\bibfnamefont {V.}~\bibnamefont {Borka~Jovanovi\'c}},\ }\href {\doibase 10.1088/1475-7516/2018/04/050} {\bibfield  {journal} {\bibinfo  {journal} {JCAP}\ }\textbf {\bibinfo {volume} {04}},\ \bibinfo {pages} {050} (\bibinfo {year} {2018})},\ \Eprint {http://arxiv.org/abs/1801.04679} {arXiv:1801.04679 [gr-qc]} \BibitemShut {NoStop}%
\bibitem [{\citenamefont {Virbhadra}\ and\ \citenamefont {Ellis}(2000)}]{Virbhadra:1999nm}%
  \BibitemOpen
  \bibfield  {author} {\bibinfo {author} {\bibfnamefont {K.~S.}\ \bibnamefont {Virbhadra}}\ and\ \bibinfo {author} {\bibfnamefont {G.~F.~R.}\ \bibnamefont {Ellis}},\ }\href {\doibase 10.1103/PhysRevD.62.084003} {\bibfield  {journal} {\bibinfo  {journal} {Phys. Rev. D}\ }\textbf {\bibinfo {volume} {62}},\ \bibinfo {pages} {084003} (\bibinfo {year} {2000})},\ \Eprint {http://arxiv.org/abs/astro-ph/9904193} {arXiv:astro-ph/9904193} \BibitemShut {NoStop}%
\bibitem [{\citenamefont {Claudel}\ \emph {et~al.}(2001)\citenamefont {Claudel}, \citenamefont {Virbhadra},\ and\ \citenamefont {Ellis}}]{Claudel:2000yi}%
  \BibitemOpen
  \bibfield  {author} {\bibinfo {author} {\bibfnamefont {C.-M.}\ \bibnamefont {Claudel}}, \bibinfo {author} {\bibfnamefont {K.~S.}\ \bibnamefont {Virbhadra}}, \ and\ \bibinfo {author} {\bibfnamefont {G.~F.~R.}\ \bibnamefont {Ellis}},\ }\href {\doibase 10.1063/1.1308507} {\bibfield  {journal} {\bibinfo  {journal} {J. Math. Phys.}\ }\textbf {\bibinfo {volume} {42}},\ \bibinfo {pages} {818} (\bibinfo {year} {2001})},\ \Eprint {http://arxiv.org/abs/gr-qc/0005050} {arXiv:gr-qc/0005050} \BibitemShut {NoStop}%
\bibitem [{\citenamefont {Virbhadra}\ and\ \citenamefont {Keeton}(2008)}]{Virbhadra:2007kw}%
  \BibitemOpen
  \bibfield  {author} {\bibinfo {author} {\bibfnamefont {K.~S.}\ \bibnamefont {Virbhadra}}\ and\ \bibinfo {author} {\bibfnamefont {C.~R.}\ \bibnamefont {Keeton}},\ }\href {\doibase 10.1103/PhysRevD.77.124014} {\bibfield  {journal} {\bibinfo  {journal} {Phys. Rev. D}\ }\textbf {\bibinfo {volume} {77}},\ \bibinfo {pages} {124014} (\bibinfo {year} {2008})},\ \Eprint {http://arxiv.org/abs/0710.2333} {arXiv:0710.2333 [gr-qc]} \BibitemShut {NoStop}%
\bibitem [{\citenamefont {Virbhadra}(2022)}]{Virbhadra:2022iiy}%
  \BibitemOpen
  \bibfield  {author} {\bibinfo {author} {\bibfnamefont {K.~S.}\ \bibnamefont {Virbhadra}},\ }\href {\doibase 10.1103/PhysRevD.106.064038} {\bibfield  {journal} {\bibinfo  {journal} {Phys. Rev. D}\ }\textbf {\bibinfo {volume} {106}},\ \bibinfo {pages} {064038} (\bibinfo {year} {2022})},\ \Eprint {http://arxiv.org/abs/2204.01879} {arXiv:2204.01879 [gr-qc]} \BibitemShut {NoStop}%
\bibitem [{\citenamefont {Joshi}(1997)}]{Joshi:1997de}%
  \BibitemOpen
  \bibfield  {author} {\bibinfo {author} {\bibfnamefont {P.~S.}\ \bibnamefont {Joshi}},\ }\href@noop {} {\  (\bibinfo {year} {1997})},\ \Eprint {http://arxiv.org/abs/gr-qc/9702036} {arXiv:gr-qc/9702036} \BibitemShut {NoStop}%
\bibitem [{\citenamefont {Joshi}(2012)}]{Joshi:2008zz}%
  \BibitemOpen
  \bibinfo {editor} {\bibfnamefont {P.~S.}\ \bibnamefont {Joshi}},\ ed.,\ \href {\doibase 10.1017/CBO9780511536274} {\emph {\bibinfo {title} {{Gravitational Collapse and Spacetime Singularities}}}},\ Cambridge Monographs on Mathematical Physics\ (\bibinfo  {publisher} {Cambridge University Press},\ \bibinfo {year} {2012})\BibitemShut {NoStop}%
\bibitem [{\citenamefont {Joshi}(2014)}]{Joshi:2013xoa}%
  \BibitemOpen
  \bibfield  {author} {\bibinfo {author} {\bibfnamefont {P.~S.}\ \bibnamefont {Joshi}},\ }\enquote {\bibinfo {title} {{Spacetime Singularities}},}\ in\ \href {\doibase 10.1007/978-3-642-41992-8_20} {\emph {\bibinfo {booktitle} {{Springer Handbook of Spacetime}}}},\ \bibinfo {editor} {edited by\ \bibinfo {editor} {\bibfnamefont {A.}~\bibnamefont {Ashtekar}}\ and\ \bibinfo {editor} {\bibfnamefont {V.}~\bibnamefont {Petkov}}}\ (\bibinfo {year} {2014})\ pp.\ \bibinfo {pages} {409--436},\ \Eprint {http://arxiv.org/abs/1311.0449} {arXiv:1311.0449 [gr-qc]} \BibitemShut {NoStop}%
\bibitem [{\citenamefont {Dey}\ \emph {et~al.}(2019)\citenamefont {Dey}, \citenamefont {Joshi}, \citenamefont {Joshi},\ and\ \citenamefont {Bambhaniya}}]{Dey:2019fpv}%
  \BibitemOpen
  \bibfield  {author} {\bibinfo {author} {\bibfnamefont {D.}~\bibnamefont {Dey}}, \bibinfo {author} {\bibfnamefont {P.~S.}\ \bibnamefont {Joshi}}, \bibinfo {author} {\bibfnamefont {A.}~\bibnamefont {Joshi}}, \ and\ \bibinfo {author} {\bibfnamefont {P.}~\bibnamefont {Bambhaniya}},\ }\href {\doibase 10.1142/S0218271819300246} {\bibfield  {journal} {\bibinfo  {journal} {Int. J. Mod. Phys. D}\ }\textbf {\bibinfo {volume} {28}},\ \bibinfo {pages} {1930024} (\bibinfo {year} {2019})},\ \Eprint {http://arxiv.org/abs/2101.06001} {arXiv:2101.06001 [gr-qc]} \BibitemShut {NoStop}%
\bibitem [{\citenamefont {Vaidya}(1951)}]{Vaidya:1951zz}%
  \BibitemOpen
  \bibfield  {author} {\bibinfo {author} {\bibfnamefont {P.}~\bibnamefont {Vaidya}},\ }\href@noop {} {\bibfield  {journal} {\bibinfo  {journal} {Proc. Natl. Inst. Sci. India A}\ }\textbf {\bibinfo {volume} {33}},\ \bibinfo {pages} {264} (\bibinfo {year} {1951})}\BibitemShut {NoStop}%
\bibitem [{\citenamefont {Penrose}(1965)}]{Penrose:1964wq}%
  \BibitemOpen
  \bibfield  {author} {\bibinfo {author} {\bibfnamefont {R.}~\bibnamefont {Penrose}},\ }\href {\doibase 10.1103/PhysRevLett.14.57} {\bibfield  {journal} {\bibinfo  {journal} {Phys. Rev. Lett.}\ }\textbf {\bibinfo {volume} {14}},\ \bibinfo {pages} {57} (\bibinfo {year} {1965})}\BibitemShut {NoStop}%
\bibitem [{\citenamefont {Hawking}\ and\ \citenamefont {Penrose}(1970)}]{Hawking:1970zqf}%
  \BibitemOpen
  \bibfield  {author} {\bibinfo {author} {\bibfnamefont {S.~W.}\ \bibnamefont {Hawking}}\ and\ \bibinfo {author} {\bibfnamefont {R.}~\bibnamefont {Penrose}},\ }\href {\doibase 10.1098/rspa.1970.0021} {\bibfield  {journal} {\bibinfo  {journal} {Proc. Roy. Soc. Lond. A}\ }\textbf {\bibinfo {volume} {314}},\ \bibinfo {pages} {529} (\bibinfo {year} {1970})}\BibitemShut {NoStop}%
\bibitem [{\citenamefont {Vertogradov}(2018)}]{Vertogradov:2018ora}%
  \BibitemOpen
  \bibfield  {author} {\bibinfo {author} {\bibfnamefont {V.}~\bibnamefont {Vertogradov}},\ }\href {\doibase 10.1142/S0217751X18501026} {\bibfield  {journal} {\bibinfo  {journal} {Int. J. Mod. Phys. A}\ }\textbf {\bibinfo {volume} {33}},\ \bibinfo {pages} {1850102} (\bibinfo {year} {2018})},\ \Eprint {http://arxiv.org/abs/2210.16131} {arXiv:2210.16131 [gr-qc]} \BibitemShut {NoStop}%
\bibitem [{\citenamefont {Shaikh}\ and\ \citenamefont {Joshi}(2019)}]{Shaikh:2019hbm}%
  \BibitemOpen
  \bibfield  {author} {\bibinfo {author} {\bibfnamefont {R.}~\bibnamefont {Shaikh}}\ and\ \bibinfo {author} {\bibfnamefont {P.~S.}\ \bibnamefont {Joshi}},\ }\href {\doibase 10.1088/1475-7516/2019/10/064} {\bibfield  {journal} {\bibinfo  {journal} {JCAP}\ }\textbf {\bibinfo {volume} {10}},\ \bibinfo {pages} {064} (\bibinfo {year} {2019})},\ \Eprint {http://arxiv.org/abs/1909.10322} {arXiv:1909.10322 [gr-qc]} \BibitemShut {NoStop}%
\bibitem [{\citenamefont {Firouzjaee}(2023)}]{Firouzjaee:2021mwl}%
  \BibitemOpen
  \bibfield  {author} {\bibinfo {author} {\bibfnamefont {J.~T.}\ \bibnamefont {Firouzjaee}},\ }\href {\doibase 10.1007/s10714-023-03073-z} {\bibfield  {journal} {\bibinfo  {journal} {Gen. Rel. Grav.}\ }\textbf {\bibinfo {volume} {55}},\ \bibinfo {pages} {38} (\bibinfo {year} {2023})},\ \Eprint {http://arxiv.org/abs/2108.10234} {arXiv:2108.10234 [gr-qc]} \BibitemShut {NoStop}%
\bibitem [{\citenamefont {Vertogradov}(2022)}]{Vertogradov:2022zuo}%
  \BibitemOpen
  \bibfield  {author} {\bibinfo {author} {\bibfnamefont {V.}~\bibnamefont {Vertogradov}},\ }\href {\doibase 10.1142/S0217751X22501858} {\bibfield  {journal} {\bibinfo  {journal} {Int. J. Mod. Phys. A}\ }\textbf {\bibinfo {volume} {37}},\ \bibinfo {pages} {2250185} (\bibinfo {year} {2022})},\ \Eprint {http://arxiv.org/abs/2209.10953} {arXiv:2209.10953 [gr-qc]} \BibitemShut {NoStop}%
\bibitem [{\citenamefont {Heydarzade}\ and\ \citenamefont {Vertogradov}(2024)}]{Heydarzade:2023gmd}%
  \BibitemOpen
  \bibfield  {author} {\bibinfo {author} {\bibfnamefont {Y.}~\bibnamefont {Heydarzade}}\ and\ \bibinfo {author} {\bibfnamefont {V.}~\bibnamefont {Vertogradov}},\ }\href {\doibase 10.1140/epjc/s10052-024-12945-w} {\bibfield  {journal} {\bibinfo  {journal} {Eur. Phys. J. C}\ }\textbf {\bibinfo {volume} {84}},\ \bibinfo {pages} {582} (\bibinfo {year} {2024})},\ \Eprint {http://arxiv.org/abs/2311.08930} {arXiv:2311.08930 [gr-qc]} \BibitemShut {NoStop}%
\bibitem [{\citenamefont {Vertogradov}(2024)}]{Vertogradov:2023uav}%
  \BibitemOpen
  \bibfield  {author} {\bibinfo {author} {\bibfnamefont {V.}~\bibnamefont {Vertogradov}},\ }\href {\doibase 10.1007/s10714-024-03244-6} {\bibfield  {journal} {\bibinfo  {journal} {Gen. Rel. Grav.}\ }\textbf {\bibinfo {volume} {56}},\ \bibinfo {pages} {59} (\bibinfo {year} {2024})},\ \Eprint {http://arxiv.org/abs/2311.15671} {arXiv:2311.15671 [gr-qc]} \BibitemShut {NoStop}%
\bibitem [{\citenamefont {Sajadi}\ \emph {et~al.}(2024)\citenamefont {Sajadi}, \citenamefont {Khodadi}, \citenamefont {Luongo},\ and\ \citenamefont {Quevedo}}]{Sajadi:2023ybm}%
  \BibitemOpen
  \bibfield  {author} {\bibinfo {author} {\bibfnamefont {S.~N.}\ \bibnamefont {Sajadi}}, \bibinfo {author} {\bibfnamefont {M.}~\bibnamefont {Khodadi}}, \bibinfo {author} {\bibfnamefont {O.}~\bibnamefont {Luongo}}, \ and\ \bibinfo {author} {\bibfnamefont {H.}~\bibnamefont {Quevedo}},\ }\href {\doibase 10.1016/j.dark.2024.101525} {\bibfield  {journal} {\bibinfo  {journal} {Phys. Dark Univ.}\ }\textbf {\bibinfo {volume} {45}},\ \bibinfo {pages} {101525} (\bibinfo {year} {2024})},\ \Eprint {http://arxiv.org/abs/2312.16081} {arXiv:2312.16081 [gr-qc]} \BibitemShut {NoStop}%
\bibitem [{\citenamefont {Hayward}(2006)}]{Hayward:2005gi}%
  \BibitemOpen
  \bibfield  {author} {\bibinfo {author} {\bibfnamefont {S.~A.}\ \bibnamefont {Hayward}},\ }\href {\doibase 10.1103/PhysRevLett.96.031103} {\bibfield  {journal} {\bibinfo  {journal} {Phys. Rev. Lett.}\ }\textbf {\bibinfo {volume} {96}},\ \bibinfo {pages} {031103} (\bibinfo {year} {2006})},\ \Eprint {http://arxiv.org/abs/gr-qc/0506126} {arXiv:gr-qc/0506126} \BibitemShut {NoStop}%
\bibitem [{\citenamefont {Petrov}(2023)}]{Petrov:2023otl}%
  \BibitemOpen
  \bibfield  {author} {\bibinfo {author} {\bibfnamefont {A.~N.}\ \bibnamefont {Petrov}},\ }\href {\doibase 10.1140/epjp/s13360-023-04514-z} {\bibfield  {journal} {\bibinfo  {journal} {Eur. Phys. J. Plus}\ }\textbf {\bibinfo {volume} {138}},\ \bibinfo {pages} {879} (\bibinfo {year} {2023})},\ \Eprint {http://arxiv.org/abs/2305.11705} {arXiv:2305.11705 [gr-qc]} \BibitemShut {NoStop}%
\bibitem [{\citenamefont {Cai}\ \emph {et~al.}(2008)\citenamefont {Cai}, \citenamefont {Cao}, \citenamefont {Hu},\ and\ \citenamefont {Kim}}]{Cai:2008mh}%
  \BibitemOpen
  \bibfield  {author} {\bibinfo {author} {\bibfnamefont {R.-G.}\ \bibnamefont {Cai}}, \bibinfo {author} {\bibfnamefont {L.-M.}\ \bibnamefont {Cao}}, \bibinfo {author} {\bibfnamefont {Y.-P.}\ \bibnamefont {Hu}}, \ and\ \bibinfo {author} {\bibfnamefont {S.~P.}\ \bibnamefont {Kim}},\ }\href {\doibase 10.1103/PhysRevD.78.124012} {\bibfield  {journal} {\bibinfo  {journal} {Phys. Rev. D}\ }\textbf {\bibinfo {volume} {78}},\ \bibinfo {pages} {124012} (\bibinfo {year} {2008})},\ \Eprint {http://arxiv.org/abs/0810.2610} {arXiv:0810.2610 [hep-th]} \BibitemShut {NoStop}%
\bibitem [{\citenamefont {Culetu}(2022)}]{Culetu:2022otf}%
  \BibitemOpen
  \bibfield  {author} {\bibinfo {author} {\bibfnamefont {H.}~\bibnamefont {Culetu}},\ }\href {\doibase 10.1142/S0218271822501243} {\bibfield  {journal} {\bibinfo  {journal} {Int. J. Mod. Phys. D}\ }\textbf {\bibinfo {volume} {31}},\ \bibinfo {pages} {2250124} (\bibinfo {year} {2022})},\ \Eprint {http://arxiv.org/abs/2202.03426} {arXiv:2202.03426 [gr-qc]} \BibitemShut {NoStop}%
\bibitem [{\citenamefont {Mann}\ \emph {et~al.}(2022)\citenamefont {Mann}, \citenamefont {Murk},\ and\ \citenamefont {Terno}}]{Mann:2021mnc}%
  \BibitemOpen
  \bibfield  {author} {\bibinfo {author} {\bibfnamefont {R.~B.}\ \bibnamefont {Mann}}, \bibinfo {author} {\bibfnamefont {S.}~\bibnamefont {Murk}}, \ and\ \bibinfo {author} {\bibfnamefont {D.~R.}\ \bibnamefont {Terno}},\ }\href {\doibase 10.1142/S0218271822300154} {\bibfield  {journal} {\bibinfo  {journal} {Int. J. Mod. Phys. D}\ }\textbf {\bibinfo {volume} {31}},\ \bibinfo {pages} {2230015} (\bibinfo {year} {2022})},\ \Eprint {http://arxiv.org/abs/2112.06515} {arXiv:2112.06515 [gr-qc]} \BibitemShut {NoStop}%
\bibitem [{\citenamefont {Simpson}\ \emph {et~al.}(2019)\citenamefont {Simpson}, \citenamefont {Martin-Moruno},\ and\ \citenamefont {Visser}}]{Simpson:2019cer}%
  \BibitemOpen
  \bibfield  {author} {\bibinfo {author} {\bibfnamefont {A.}~\bibnamefont {Simpson}}, \bibinfo {author} {\bibfnamefont {P.}~\bibnamefont {Martin-Moruno}}, \ and\ \bibinfo {author} {\bibfnamefont {M.}~\bibnamefont {Visser}},\ }\href {\doibase 10.1088/1361-6382/ab28a5} {\bibfield  {journal} {\bibinfo  {journal} {Class. Quant. Grav.}\ }\textbf {\bibinfo {volume} {36}},\ \bibinfo {pages} {145007} (\bibinfo {year} {2019})},\ \Eprint {http://arxiv.org/abs/1902.04232} {arXiv:1902.04232 [gr-qc]} \BibitemShut {NoStop}%
\bibitem [{\citenamefont {Baccetti}\ \emph {et~al.}(2019)\citenamefont {Baccetti}, \citenamefont {Murk},\ and\ \citenamefont {Terno}}]{Baccetti:2018qrp}%
  \BibitemOpen
  \bibfield  {author} {\bibinfo {author} {\bibfnamefont {V.}~\bibnamefont {Baccetti}}, \bibinfo {author} {\bibfnamefont {S.}~\bibnamefont {Murk}}, \ and\ \bibinfo {author} {\bibfnamefont {D.~R.}\ \bibnamefont {Terno}},\ }\href {\doibase 10.1103/PhysRevD.100.064054} {\bibfield  {journal} {\bibinfo  {journal} {Phys. Rev. D}\ }\textbf {\bibinfo {volume} {100}},\ \bibinfo {pages} {064054} (\bibinfo {year} {2019})},\ \Eprint {http://arxiv.org/abs/1812.07727} {arXiv:1812.07727 [gr-qc]} \BibitemShut {NoStop}%
\bibitem [{\citenamefont {Hossenfelder}\ \emph {et~al.}(2010)\citenamefont {Hossenfelder}, \citenamefont {Modesto},\ and\ \citenamefont {Premont-Schwarz}}]{Hossenfelder:2009fc}%
  \BibitemOpen
  \bibfield  {author} {\bibinfo {author} {\bibfnamefont {S.}~\bibnamefont {Hossenfelder}}, \bibinfo {author} {\bibfnamefont {L.}~\bibnamefont {Modesto}}, \ and\ \bibinfo {author} {\bibfnamefont {I.}~\bibnamefont {Premont-Schwarz}},\ }\href {\doibase 10.1103/PhysRevD.81.044036} {\bibfield  {journal} {\bibinfo  {journal} {Phys. Rev. D}\ }\textbf {\bibinfo {volume} {81}},\ \bibinfo {pages} {044036} (\bibinfo {year} {2010})},\ \Eprint {http://arxiv.org/abs/0912.1823} {arXiv:0912.1823 [gr-qc]} \BibitemShut {NoStop}%
\bibitem [{\citenamefont {Ghosh}\ and\ \citenamefont {Saraykar}(2000)}]{Ghosh:2000bc}%
  \BibitemOpen
  \bibfield  {author} {\bibinfo {author} {\bibfnamefont {S.~G.}\ \bibnamefont {Ghosh}}\ and\ \bibinfo {author} {\bibfnamefont {R.~V.}\ \bibnamefont {Saraykar}},\ }\href {\doibase 10.1103/PhysRevD.62.107502} {\bibfield  {journal} {\bibinfo  {journal} {Phys. Rev. D}\ }\textbf {\bibinfo {volume} {62}},\ \bibinfo {pages} {107502} (\bibinfo {year} {2000})},\ \Eprint {http://arxiv.org/abs/gr-qc/0111080} {arXiv:gr-qc/0111080} \BibitemShut {NoStop}%
\bibitem [{\citenamefont {Ghosh}(2000)}]{Ghosh:2000ud}%
  \BibitemOpen
  \bibfield  {author} {\bibinfo {author} {\bibfnamefont {S.~G.}\ \bibnamefont {Ghosh}},\ }\href {\doibase 10.1103/PhysRevD.62.127505} {\bibfield  {journal} {\bibinfo  {journal} {Phys. Rev. D}\ }\textbf {\bibinfo {volume} {62}},\ \bibinfo {pages} {127505} (\bibinfo {year} {2000})},\ \Eprint {http://arxiv.org/abs/gr-qc/0106060} {arXiv:gr-qc/0106060} \BibitemShut {NoStop}%
\bibitem [{\citenamefont {Ghosh}\ and\ \citenamefont {Deshkar}(2008)}]{Ghosh:2008jca}%
  \BibitemOpen
  \bibfield  {author} {\bibinfo {author} {\bibfnamefont {S.~G.}\ \bibnamefont {Ghosh}}\ and\ \bibinfo {author} {\bibfnamefont {D.~W.}\ \bibnamefont {Deshkar}},\ }\href {\doibase 10.1103/PhysRevD.77.047504} {\bibfield  {journal} {\bibinfo  {journal} {Phys. Rev. D}\ }\textbf {\bibinfo {volume} {77}},\ \bibinfo {pages} {047504} (\bibinfo {year} {2008})},\ \Eprint {http://arxiv.org/abs/0801.2710} {arXiv:0801.2710 [gr-qc]} \BibitemShut {NoStop}%
\bibitem [{\citenamefont {Ghosh}(2015)}]{Ghosh:2014pba}%
  \BibitemOpen
  \bibfield  {author} {\bibinfo {author} {\bibfnamefont {S.~G.}\ \bibnamefont {Ghosh}},\ }\href {\doibase 10.1140/epjc/s10052-015-3740-y} {\bibfield  {journal} {\bibinfo  {journal} {Eur. Phys. J. C}\ }\textbf {\bibinfo {volume} {75}},\ \bibinfo {pages} {532} (\bibinfo {year} {2015})},\ \Eprint {http://arxiv.org/abs/1408.5668} {arXiv:1408.5668 [gr-qc]} \BibitemShut {NoStop}%
\bibitem [{\citenamefont {Nasereldin}\ and\ \citenamefont {Lake}(2023)}]{Nasereldin:2023qph}%
  \BibitemOpen
  \bibfield  {author} {\bibinfo {author} {\bibfnamefont {S.}~\bibnamefont {Nasereldin}}\ and\ \bibinfo {author} {\bibfnamefont {K.}~\bibnamefont {Lake}},\ }\href {\doibase 10.1103/PhysRevD.108.124064} {\bibfield  {journal} {\bibinfo  {journal} {Phys. Rev. D}\ }\textbf {\bibinfo {volume} {108}},\ \bibinfo {pages} {124064} (\bibinfo {year} {2023})},\ \Eprint {http://arxiv.org/abs/2307.06139} {arXiv:2307.06139 [gr-qc]} \BibitemShut {NoStop}%
\bibitem [{\citenamefont {Sharif}\ and\ \citenamefont {Yousaf}(2016)}]{Sharif:2015vya}%
  \BibitemOpen
  \bibfield  {author} {\bibinfo {author} {\bibfnamefont {M.}~\bibnamefont {Sharif}}\ and\ \bibinfo {author} {\bibfnamefont {Z.}~\bibnamefont {Yousaf}},\ }\href {\doibase 10.1007/s10773-015-2681-4} {\bibfield  {journal} {\bibinfo  {journal} {Int. J. Theor. Phys.}\ }\textbf {\bibinfo {volume} {55}},\ \bibinfo {pages} {470} (\bibinfo {year} {2016})}\BibitemShut {NoStop}%
\bibitem [{\citenamefont {Berezin}\ \emph {et~al.}(2016)\citenamefont {Berezin}, \citenamefont {Dokuchaev},\ and\ \citenamefont {Eroshenko}}]{Berezin:2016ubu}%
  \BibitemOpen
  \bibfield  {author} {\bibinfo {author} {\bibfnamefont {V.~A.}\ \bibnamefont {Berezin}}, \bibinfo {author} {\bibfnamefont {V.~I.}\ \bibnamefont {Dokuchaev}}, \ and\ \bibinfo {author} {\bibfnamefont {Y.~N.}\ \bibnamefont {Eroshenko}},\ }\href {\doibase 10.1088/0264-9381/33/14/145003} {\bibfield  {journal} {\bibinfo  {journal} {Class. Quant. Grav.}\ }\textbf {\bibinfo {volume} {33}},\ \bibinfo {pages} {145003} (\bibinfo {year} {2016})},\ \Eprint {http://arxiv.org/abs/1603.00849} {arXiv:1603.00849 [gr-qc]} \BibitemShut {NoStop}%
\bibitem [{\citenamefont {Babichev}\ \emph {et~al.}(2012)\citenamefont {Babichev}, \citenamefont {Dokuchaev},\ and\ \citenamefont {Eroshenko}}]{Babichev:2012sg}%
  \BibitemOpen
  \bibfield  {author} {\bibinfo {author} {\bibfnamefont {E.}~\bibnamefont {Babichev}}, \bibinfo {author} {\bibfnamefont {V.}~\bibnamefont {Dokuchaev}}, \ and\ \bibinfo {author} {\bibfnamefont {Y.}~\bibnamefont {Eroshenko}},\ }\href {\doibase 10.1088/0264-9381/29/11/115002} {\bibfield  {journal} {\bibinfo  {journal} {Class. Quant. Grav.}\ }\textbf {\bibinfo {volume} {29}},\ \bibinfo {pages} {115002} (\bibinfo {year} {2012})},\ \Eprint {http://arxiv.org/abs/1202.2836} {arXiv:1202.2836 [gr-qc]} \BibitemShut {NoStop}%
\bibitem [{\citenamefont {Malafarina}(2016)}]{Malafarina:2016yuf}%
  \BibitemOpen
  \bibfield  {author} {\bibinfo {author} {\bibfnamefont {D.}~\bibnamefont {Malafarina}},\ }\href {\doibase 10.1103/PhysRevD.93.104042} {\bibfield  {journal} {\bibinfo  {journal} {Phys. Rev. D}\ }\textbf {\bibinfo {volume} {93}},\ \bibinfo {pages} {104042} (\bibinfo {year} {2016})},\ \Eprint {http://arxiv.org/abs/1605.03312} {arXiv:1605.03312 [gr-qc]} \BibitemShut {NoStop}%
\bibitem [{\citenamefont {Harko}(2003)}]{Harko:2003hs}%
  \BibitemOpen
  \bibfield  {author} {\bibinfo {author} {\bibfnamefont {T.}~\bibnamefont {Harko}},\ }\href {\doibase 10.1103/PhysRevD.68.064005} {\bibfield  {journal} {\bibinfo  {journal} {Phys. Rev. D}\ }\textbf {\bibinfo {volume} {68}},\ \bibinfo {pages} {064005} (\bibinfo {year} {2003})},\ \Eprint {http://arxiv.org/abs/gr-qc/0307064} {arXiv:gr-qc/0307064} \BibitemShut {NoStop}%
\bibitem [{\citenamefont {Vertogradov}\ and\ \citenamefont {\"Ovg\"un}(2024{\natexlab{a}})}]{Vertogradov:2024qpf}%
  \BibitemOpen
  \bibfield  {author} {\bibinfo {author} {\bibfnamefont {V.}~\bibnamefont {Vertogradov}}\ and\ \bibinfo {author} {\bibfnamefont {A.}~\bibnamefont {\"Ovg\"un}},\ }\href {\doibase 10.1016/j.dark.2024.101541} {\bibfield  {journal} {\bibinfo  {journal} {Phys. Dark Univ.}\ }\textbf {\bibinfo {volume} {45}},\ \bibinfo {pages} {101541} (\bibinfo {year} {2024}{\natexlab{a}})},\ \Eprint {http://arxiv.org/abs/2404.04046} {arXiv:2404.04046 [gr-qc]} \BibitemShut {NoStop}%
\bibitem [{\citenamefont {Vagnozzi}\ \emph {et~al.}(2023)\citenamefont {Vagnozzi} \emph {et~al.}}]{Vagnozzi:2022moj}%
  \BibitemOpen
  \bibfield  {author} {\bibinfo {author} {\bibfnamefont {S.}~\bibnamefont {Vagnozzi}} \emph {et~al.},\ }\href {\doibase 10.1088/1361-6382/acd97b} {\bibfield  {journal} {\bibinfo  {journal} {Class. Quant. Grav.}\ }\textbf {\bibinfo {volume} {40}},\ \bibinfo {pages} {165007} (\bibinfo {year} {2023})},\ \Eprint {http://arxiv.org/abs/2205.07787} {arXiv:2205.07787 [gr-qc]} \BibitemShut {NoStop}%
\bibitem [{\citenamefont {Mkenyeleye}\ \emph {et~al.}(2014)\citenamefont {Mkenyeleye}, \citenamefont {Goswami},\ and\ \citenamefont {Maharaj}}]{Mkenyeleye:2014dwa}%
  \BibitemOpen
  \bibfield  {author} {\bibinfo {author} {\bibfnamefont {M.~D.}\ \bibnamefont {Mkenyeleye}}, \bibinfo {author} {\bibfnamefont {R.}~\bibnamefont {Goswami}}, \ and\ \bibinfo {author} {\bibfnamefont {S.~D.}\ \bibnamefont {Maharaj}},\ }\href {\doibase 10.1103/PhysRevD.90.064034} {\bibfield  {journal} {\bibinfo  {journal} {Phys. Rev. D}\ }\textbf {\bibinfo {volume} {90}},\ \bibinfo {pages} {064034} (\bibinfo {year} {2014})},\ \Eprint {http://arxiv.org/abs/1407.4309} {arXiv:1407.4309 [gr-qc]} \BibitemShut {NoStop}%
\bibitem [{\citenamefont {Vertogradov}\ and\ \citenamefont {\"Ovg\"un}(2024{\natexlab{b}})}]{Vertogradov:2024dpa}%
  \BibitemOpen
  \bibfield  {author} {\bibinfo {author} {\bibfnamefont {V.}~\bibnamefont {Vertogradov}}\ and\ \bibinfo {author} {\bibfnamefont {A.}~\bibnamefont {\"Ovg\"un}},\ }\href {\doibase 10.1016/j.physletb.2024.138758} {\bibfield  {journal} {\bibinfo  {journal} {Phys. Lett. B}\ }\textbf {\bibinfo {volume} {854}},\ \bibinfo {pages} {138758} (\bibinfo {year} {2024}{\natexlab{b}})},\ \Eprint {http://arxiv.org/abs/2404.18536} {arXiv:2404.18536 [gr-qc]} \BibitemShut {NoStop}%
\end{thebibliography}%

\end{document}